\documentclass[iop,numberedappendix]{emulateapj}

\usepackage{multirow}
\usepackage{hyperref}

\newcommand{\kms}{\mbox{\,km$\,$s$^{-1}$}}
\newcommand{\K}{\mbox{\,K}}
\newcommand{\msun}{\mbox{$\rm M_{\sun}$}}
\newcommand{\cmd}{\mbox{cm$^{-2}$}}
\newcommand{\nlq}{}
\newcommand{\nlqb}{}

\received{October 23,2019}
\accepted{January 15, 2020}

\shorttitle{Large-scale molecular gas distribution in the M17}

\begin{document}

\title{Large-scale molecular gas distribution in the M17 cloud complex:\\ dense gas conditions of massive star formation?}

\author{Quang Nguyen-Luong}
\affiliation{McMaster University, 1 James St N, Hamilton, ON, L8P 1A2, Canada}
\affiliation{Graduate School of Natural Sciences, Nagoya City University, Mizuho-ku, Nagoya, Aichi 467-8601, Japan}
\email{nguyeq12@mcmaster.ca}

\author{Fumitaka Nakamura}
\affiliation{National Astronomical Observatory of Japan, 2-21-1 Osawa, Mitaka, Tokyo 181-8588, Japan}
\affiliation{Department of Astronomy, The University of Tokyo, Hongo, Tokyo 113-0033, Japan}
\affiliation{The Graduate University for Advanced Studies (SOKENDAI), 2-21-1 Osawa, Mitaka, Tokyo 181-0015, Japan}
\author{Koji Sugitani}
\affiliation{Graduate School of Natural Sciences, Nagoya City University, Mizuho-ku, Nagoya, Aichi 467-8601, Japan}

\author{Tomomi Shimoikura}
\affiliation{Faculty of Social Information Studies, Otsuma Women's University, Chiyoda-ku,Tokyo,  102-8357, Japan}

\author{Kazuhito Dobashi}
\affiliation{Department of Astronomy and Earth Sciences, Tokyo Gakugei University, 4-1-1 Nukuikitamachi, Koganei, Tokyo 184-8501, Japan}

\author{Shinichi W. Kinoshita}
\affiliation{Department of Astronomy, The University of Tokyo, Hongo, Tokyo 113-0033, Japan}
\affiliation{National Astronomical Observatory of Japan, 2-21-1 Osawa, Mitaka, Tokyo 181-8588, Japan}

\author{Kee-Tae Kim}
\affiliation{Korea Astronomy \& Space Science Institute, 776 Daedeokdae-ro, Yuseong-gu, Daejeon 34055, Republic of Korea}
\affiliation{University of Science and Technology, Korea (UST), 217 Gajeong-ro, Yuseong-gu, Daejeon 34113, Republic of Korea}

\author{Hynwoo Kang}
\affiliation{Korea Astronomy \& Space Science Institute, 776 Daedeokdae-ro, Yuseong-gu, Daejeon 34055, Republic of Korea}

\author{Patricio Sanhueza}
\affiliation{National Astronomical Observatory of Japan, 2-21-1 Osawa, Mitaka, Tokyo 181-8588, Japan}

\author{Neal J. Evans II}
\affiliation{Korea Astronomy \& Space Science Institute, 776 Daedeokdae-ro, Yuseong-gu, Daejeon 34055, Republic of Korea}
\affiliation{Department of Astronomy, The University of Texas at Austin, 2515 Speedway, Stop C1400, Austin, TX 78712-1205, USA}

\author{Glenn J. White}
\affiliation{Department of Physics and Astronomy, The Open University, Walton Hall, Milton Keynes, MK7 6AA, UK}  \affiliation{RAL Space, STFC Rutherford Appleton Laboratory, Chilton, Didcot, Oxfordshire, OX11 0QX, UK}

\begin{abstract}
The non-uniform distribution of gas and protostars in molecular clouds is caused by combinations of various physical processes that are difficult to separate. We explore this non-uniform distribution in the M17 molecular cloud complex that hosts massive star formation activity using the $^{12}$CO ($J=1-0$) and $^{13}$CO ($J=1-0$) emission lines obtained with the Nobeyama 45m telescope. 
Differences in clump properties such as mass, size, and gravitational boundedness reflect the different evolutionary stages of the M17-H{\scriptsize II} and M17-IRDC clouds.
Clumps in the M17-H{\scriptsize II} cloud are denser, more compact, and more gravitationally bound than those in M17-IRDC. While M17-H{\scriptsize II} hosts a large fraction of very dense gas (27\%) that has column density larger than the threshold of $\sim$ 1 g cm$^{-2}$ theoretically predicted for massive star formation, this very dense gas is deficient in M17-IRDC (0.46\%). 
Our HCO$^+$ ($J=1-0$) and HCN ($J=1-0$) observations with the TRAO 14m telescope, {\nlqb trace all gas with column density higher than $3\times 10^{22}$ cm$^{-2}$}, confirm the deficiency of high density ($\gtrsim 10^5$ cm$^{-3}$) gas in M17-IRDC. Although M17-IRDC is massive enough to potentially form massive stars, its deficiency of very dense gas and gravitationally bound clumps can explain the current lack of massive star formation.

\end{abstract}

\section{Introduction} \label{sec:intro}
\begin{figure*}[htbp]
 \begin{center}
   \includegraphics[width=17.9cm]{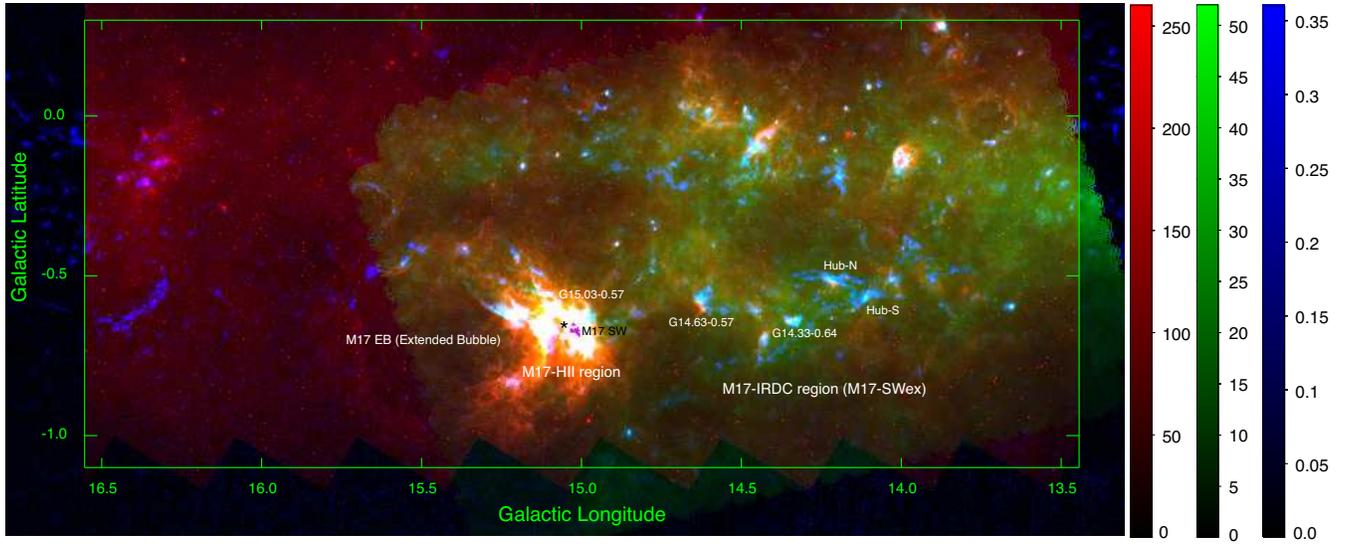} \\
   \end{center}
\caption{Three-color image of the M17 region as seen at 24\,\micron\, in MJy/sr ({\it red}, Spitzer), 250\,\micron\, in Jy/beam ({\it green}, Herschel), and 870\,\micron\, in Jy/beam ({\it blue}, APEX).}
\label{fig:m17IR}
\end{figure*}

The discovery of Carbon monoxide (CO) in the interstellar medium opened a new window into the molecular gas universe  \citep{wilson70}.  Understanding how molecular gas is organized into structures is important since it has an essential role in nurturing the star and planet formation process. Besides a degree-resolution all-sky area in CO ($J=1-0$) \citep{dame01}, numerous higher resolution wide-field CO surveys have presented three-dimensional (3D) space-space-velocity structures of the molecular gas environments in more detail (e.g. \citealt{schneider10,carlhoff13,dempsey13,barnes15}). {\nlqb Wide-field mapping of denser gas tracers, i.e., HCO$^{+}$, HCN, CS also probe into the inner dense region of the molecular clouds \citep{wu10,kauffmann17,pety17}.} 

Using one of the world's largest mm-waveband telescopes, the 45m telescope of the Nobeyama Radio Observatory (NRO 45m), we have conducted a high spatial resolution, highly sensitive, large dynamical range wide-field survey of CO  ($J=1-0$) and other gases to explore the molecular cloud structure in both low-mass and high-mass star-forming regions. The project is named ``Star Formation Legacy project" and presented in \cite{nakamura19a}. The detailed observational results for the individual regions are given in other articles (Orion A: \citealt{ishii19}, \citealt{tanabe20}; \citealt{nakamura19b}; M17: \citealt{shimoikura19a}, \citealt{sugitani19}; Aquila Rift: \citealt{shimoikura18},
\citealt{kusune19};  other regions: \citealt{dobashi19a}, \citealt{dobashi19b}).

As a part of the project, we investigate the global molecular gas distribution of the M17 region to understand the role of dense gas in star formation in M17 (see Figure \ref{fig:m17IR} for the wide-field infrared and submillimeter image of the mapped region). The $^{12}$CO ($J=1-0$) and $^{13}$CO ($J=1-0$) data from the NRO 45m telescope is complemented by HCN ($J=1-0$) and HCO$^{+}$ ($J=1-0$) observed by the Taeduk Radio Astronomy Observatory (TRAO) 14m telescope. We summarize an overview of the M17 complex in Section~\ref{sect:overview}. Section \ref{sect:observation} describes the detailed observations and data used in this paper. The global structure of molecular gas and dense gas is examined in Section\,\ref{sect:results} and the role of dense gas in star formation in M17 is discussed in Section \ref{sect:discussion}. Finally, we summarize the results in Section \ref{sect:conclusion}.

\section{Overview of the M17 region}
\label{sect:overview}

The M17 complex is a $\sim2\degr\times2\degr$ molecular cloud complex surrounding M17 nebula (also called The Omega, or The Horseshoe, or The Swan nebula) and is located in the Sagittarius spiral arm  \citep{elmegreen79,reid19}. The M17 nebula is excited by 1 Myr-old the NGC~6618 loose (radius $>$ 1 pc) open cluster which contains hundreds of stars with spectral types earlier than B9 \citep{lada91}. The MYStIX (Massive Young Star-Forming Complex Study in IR and X-ray) survey with the Chandra X-Ray Observatory counted a total of 16000 stars in NGC~6618 (M17) cluster \citep{kuhn15}. It is the second most populated cluster after the Carina cluster in the MYStIX survey \citep{kuhn15}.  For comparison, while NGC~6611 and Orion Nebula Cluster have peak stellar surface densities of around 10--100 stars pc$^{-2}$, that of NGC~6618 is remarkably much higher, $> 1000$ stars pc$^{-2}$. 
The H{\scriptsize II} region (M17-H{\scriptsize II} region) surrounding the cluster has opened a large gap at its edge which lets stellar radiation and winds escape exciting a diffuse X-Ray emitting region observed with the Chandra Observatory \citep{townsley03}. In addition to the relatively mature  H{\scriptsize II} region around the cluster, other notable star-forming regions have been discovered in the vicinity of M17 such as the immediate environment of M17 \citep{ando02}, or the M17 Infrared Dark Cloud (M17-IRDC, also known as M7-SWex) that contains the IRDC G14.225-0.506 \citep{povich10,povich16,ohashi16}. M17 forms a larger molecular cloud complex together with M16 cloud as suggested by \cite{nguyenluong16}. 

Parallax distances of $1.83^{+0.08}_{-0.07}$ kpc and  $1.98^{+0.14}_{-0.12}$ kpc by maser monitoring have been determined towards two dust clumps in M17, G014.63-00.57 and G015.03-00.57, respectively \citep{honma12,wu14}. These parallax distances are larger than photometric distances of $1.3\pm0.4$ kpc \citep{hanson97} and $1.6\pm0.3$ kpc \citep{nielbock01}, obtained by the analysis of the main-sequence OB stars. Note that \citet{chini80} derived a distance of 2.2\,kpc to M17 based on the multi-colour photometry. 
Other parallax measurements to M17-H{\scriptsize II} region have suggested a distance of 2.0 kpc \citep{xu11},  $1.9\pm 0.1$ kpc \citep{wu19}, or $2.04^{+0.16}_{-0.17}$ kpc\citep{chibueze16}. To be consistent with other papers in our project \citep{sugitani19, shimoikura19a}, 
we adopt 2~kpc to be the distance to the entire M17 complex.

\begin{figure*}[!htbp]
 \begin{center}
 \includegraphics[width=17.9cm]{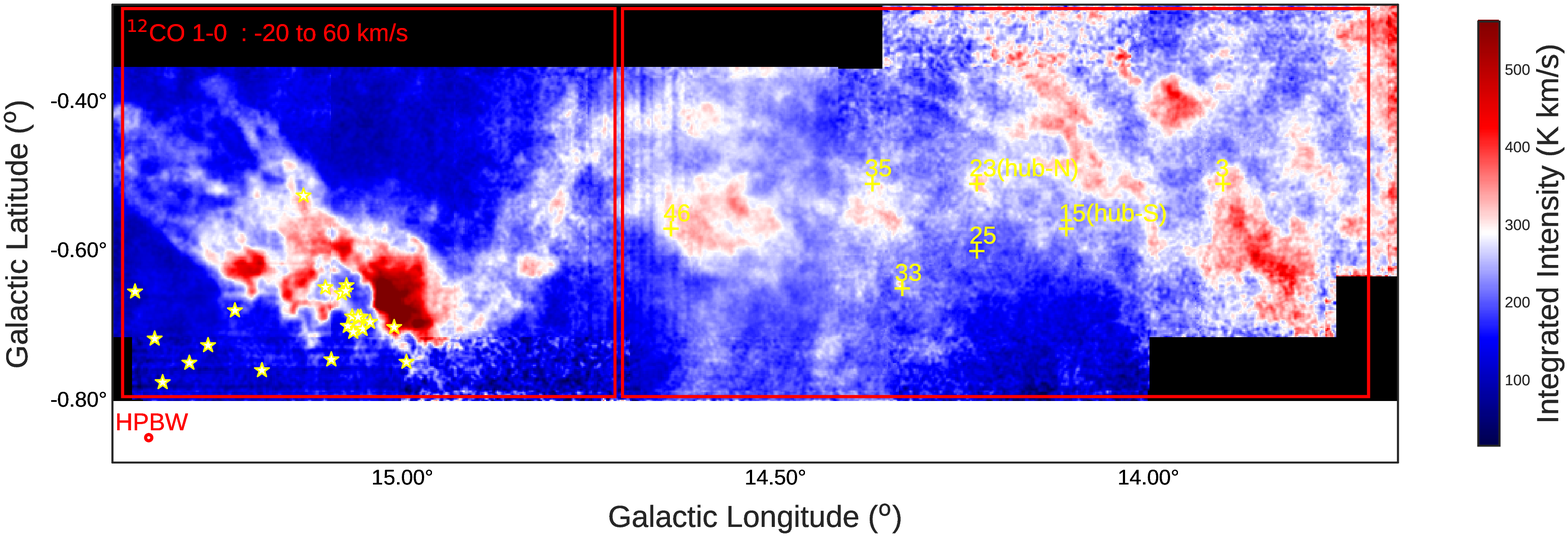}\\
 \includegraphics[width=17.9cm]{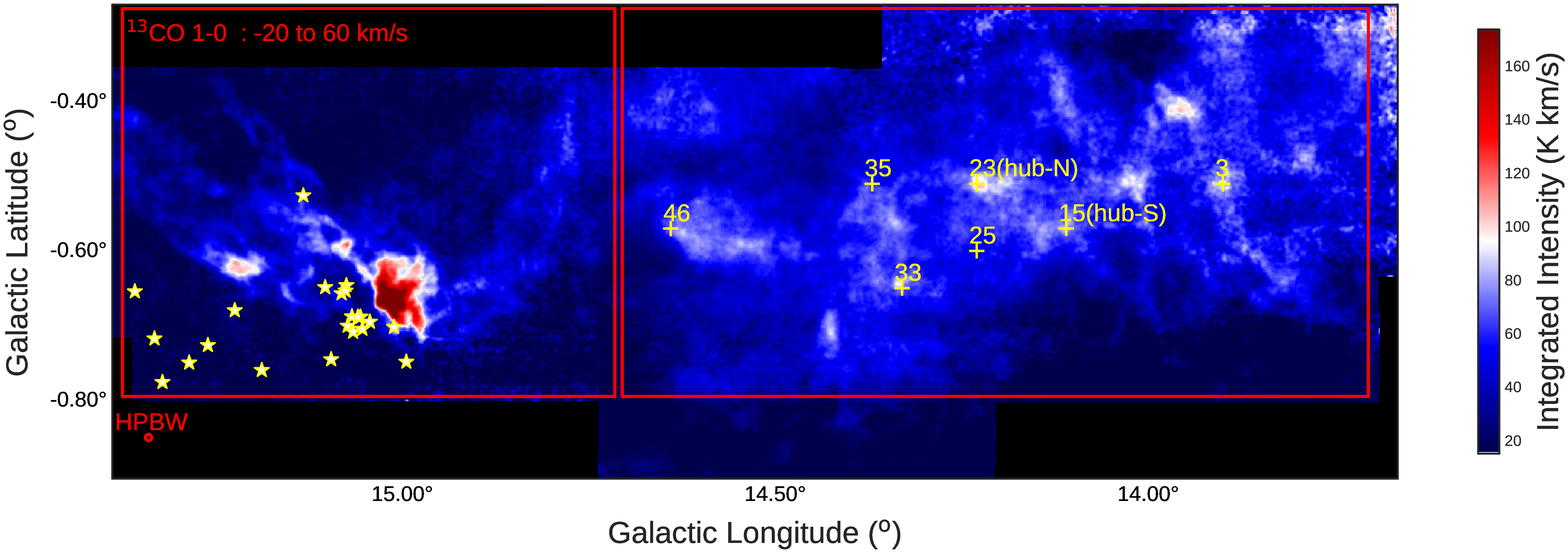}\\
  \vskip -1cm
 \end{center}
\caption{The M17 complex in {\bf (upper)} $^{12}$CO\,($J=1-0$)  and {\bf (lower)} 
$^{13}$CO\,($J=1-0$) from the NRO 45m star formation survey, integrated over the entire velocity range {(from -20 to 60\,\kms)}. The red rectangles outline the two prominent star-forming regions: M17-H{\scriptsize II} (left) and M17-IRDC (right).  {The star symbols pinpoint the OB star clusters responsible for the giant H{\scriptsize II} region.} The cross symbols pinpoint the locations of the massive cores (M $>$ 500\,\msun) reported by \cite{shimoikura19a}. 
}
\label{fig:m17_intall}
\end{figure*}

\citet{elmegreen79} found a velocity gradient in the M17 molecular cloud complex from north-east to south-west based on the low-resolution $^{12}$CO ($J=1-0$) and $^{13}$CO ($J=1-0$) observations. They suggested that the gradient
may be an outcome of the recent passage of a Galactic spiral density wave, 
which has triggered star formation in M17.
The spiral density waves compress the interstellar gas, promoting the formation of giant molecular clouds. The spiral density waves also enhance the collision rates of molecular clouds, which can trigger massive star formation and star cluster formation efficiently \citep{scoville86,tan00,nakamura12,wu17,fukui14,dobashi19b}.
In fact, \citet{nishimura18} recently found evidence for a cloud-cloud collision near the M17-H{\scriptsize II} region. However, streaming motions from the spiral waves can also inhibit massive star formation, a fact that has been seen in other galaxies such as M51 \citep{meidt13}.
In either case, the M17 complex, as a whole, is well-suited to study the effect of dynamical compression of interstellar gas by the Galactic spiral density wave. Interestingly, a second compression at the interface of the M17-H{\scriptsize II} region when the OB star clusters compress the edge of the cloud also occured which can be seen in CO ($J=3-2$) \citep{rainey87}
and also in high density tracer such as HCN ($J=4-3$) \citep{white82}.

From the Spitzer observations, \citet{povich10} and \citet{povich16} discovered that the mass function of young stellar objects (YSOs) around the M17-H{\scriptsize II} region seems consistent with the Salpeter IMF, whereas that in the M17-IRDC is significantly steeper than the Salpeter IMF. 
In other words, the high-mass stellar population in M17-IRDC is deficient.
This fact makes the M17 region an attractive test case for models of high-mass star formation.

\section{Observations}
\label{sect:observation}
\subsection{CO observations from the NRO 45m star formation project}

The data come from the NRO 45m star formation project (PI: Fumitaka Nakamura) which observed $^{12}$CO ($J=1-0$), $^{13}$CO ($J=1-0$), C$^{18}$O ($J=1-0$), N$_2$H$^+$ ($J=1-0$) and CCS ($J_N=8_7-7_6$) lines toward a sample of star-forming regions: M17, Orion, and Aquila Rift. M17 is the most distant star-forming region in the survey. 
We carried out the mapping observations toward M17 between December 2014 and March 2017. The three CO isotopologue lines 
were observed using the four-beam dual polarization, sideband-separating SIS FOREST receiver \citep{minamidani16}. However, the C$^{18}$O coverage is smaller than those of $^{12}$CO and $^{13}$CO, due to malfunction of a sub-reflector system in one period of observations. See \cite{nakamura19a} for more details of the observations.

The telescope HPBW beam size is $\sim15\arcsec$ and the main-beam efficiency $\eta_{\rm MB}$ $\sim40\%$ at 115 GHz. We used the SAM45 spectrometer as the backend which provided a bandwidth of 63 MHz and a frequency resolution of 15.26 kHz, corresponding to a velocity resolution of $\sim0.04\,\kms$. Standard on-the-fly (OTF) mapping techniques were used to carry out the mapping observations. Details on the observation procedure such as OFF-positions, submapping integration time, efficiency can be found in \cite{nakamura19a}.

\begin{figure}[htbp]
 \hspace{-0.4cm}
 \includegraphics[width=8.9cm]{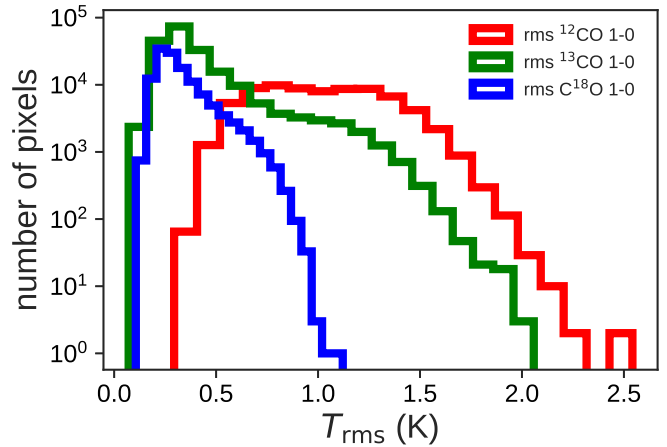}
\caption{The histograms of the rms maps of $^{12}$CO\,($J=1-0$) ({\it red}), $^{13}$CO\,($J=1-0$) ({\it green}), and C$^{18}$O\,($J=1-0$) ({\it blue}) of the M17 data.
}
\label{fig:m17_rmshist}
\end{figure}

The raw data were reduced by the NRO data reduction tool, NOSTAR. All three $^{12}$CO, $^{13}$CO, and C$^{18}$O data were convolved to $22\arcsec$\, beam size and reprojected to
a common $7.5\arcsec \times 7.5\arcsec$ grid
to facilitate our analysis. 
Figure \ref{fig:m17_intall} shows the $^{12}$CO and $^{13}$CO intensity maps integrated from --20 \kms\, to 60 \kms.
The rms noise levels 
were calculated as the average of the emission-free channels from --18 to --11\,\kms. The average rms of $^{12}$CO, $^{13}$CO, and C$^{18}$O are $\sim$ 1.0, 0.4, and 0.3\,K per 7.5\arcsec -pixel and per 0.1\,\kms channel, respectively (Figure\,\ref{fig:m17_rmshist}). The data are available online\footnote{http://jvo.nao.ac.jp/portal/v2/}. 

\begin{figure*}[!hbtp]
\vskip -0.1cm
\hspace{-0.3cm}
 $\begin{array}{ccc}
\includegraphics[width=6cm]{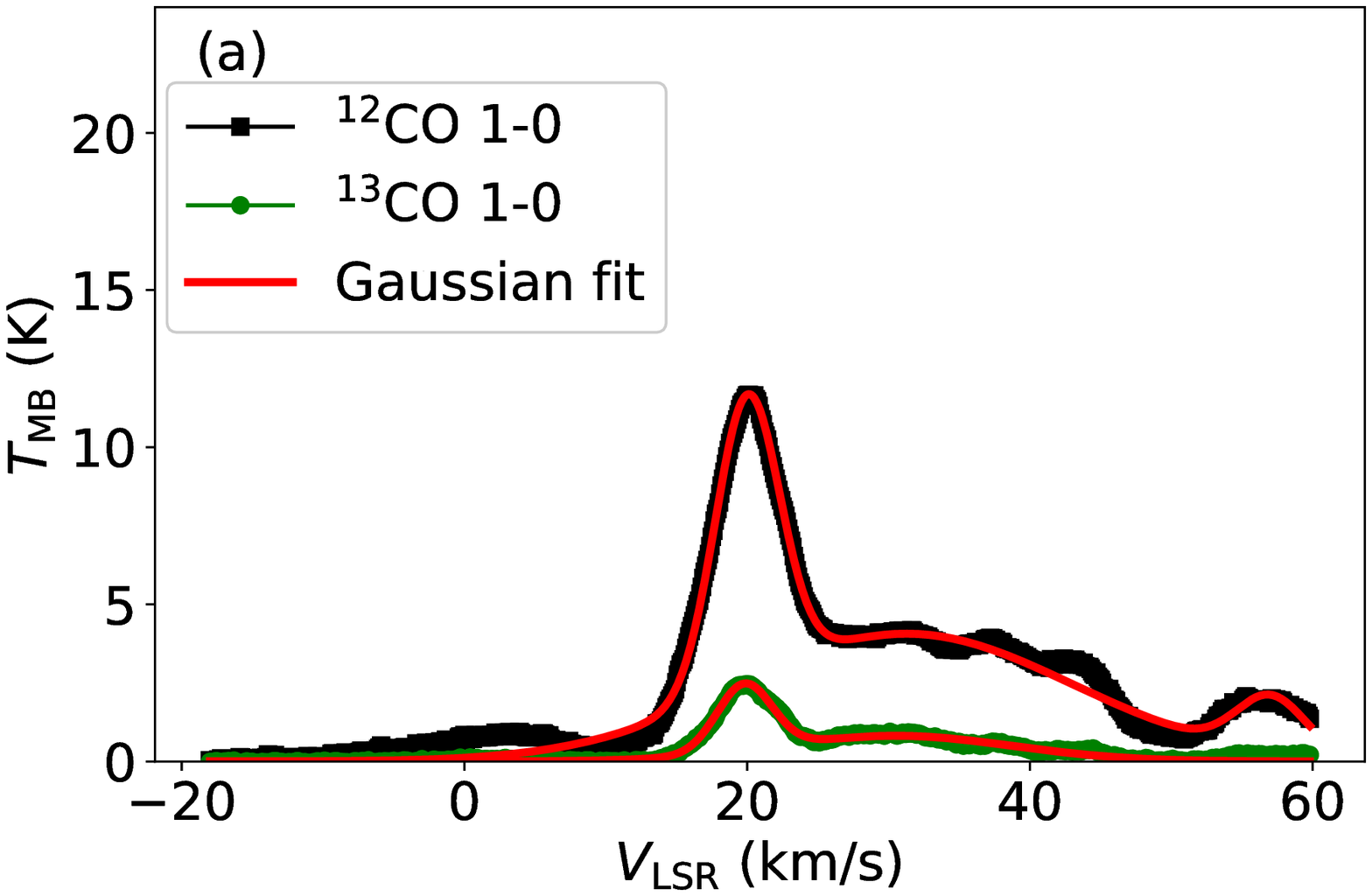} &
\hspace{-0.2cm}
\includegraphics[width=6cm]{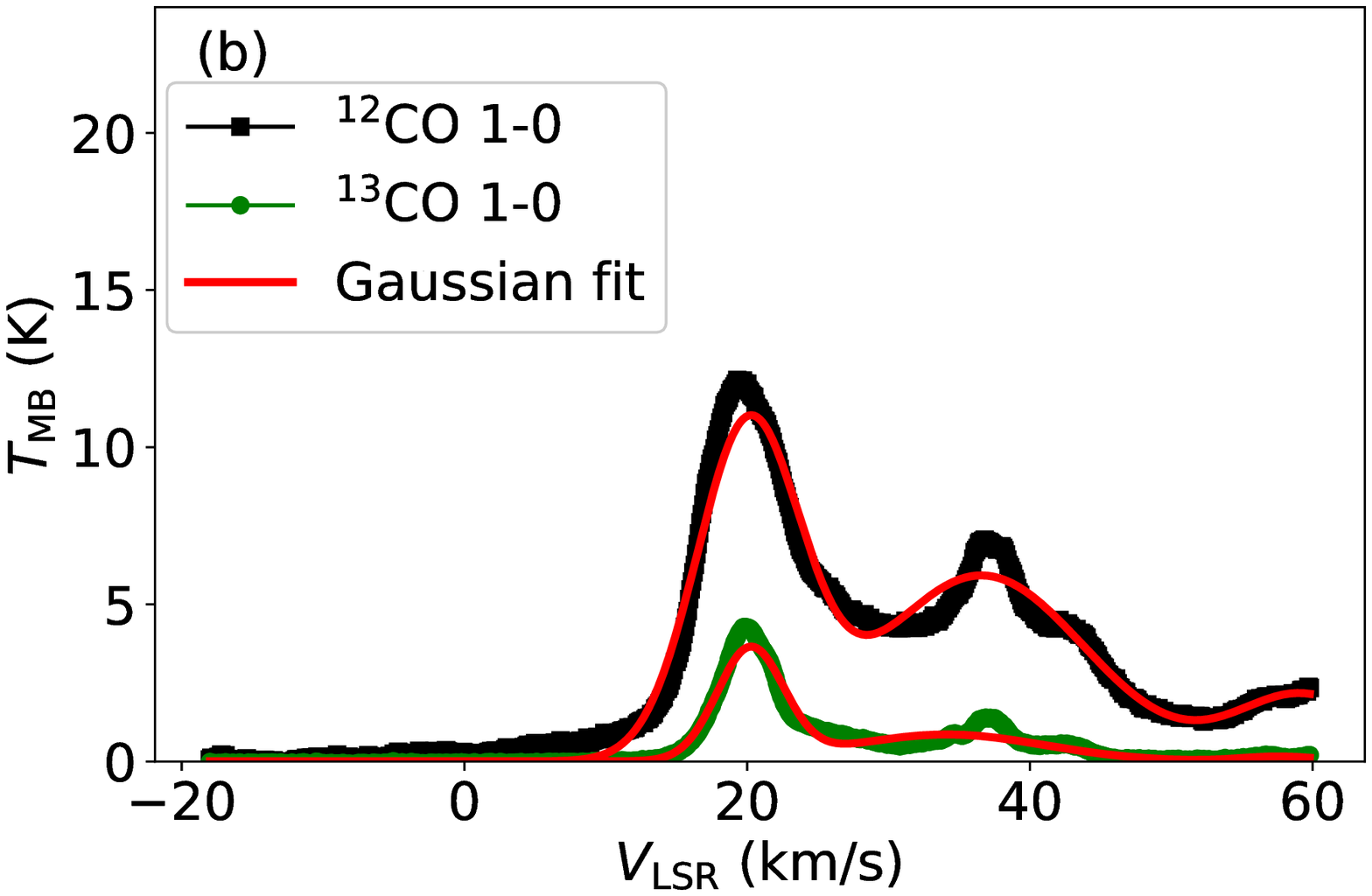} &
\includegraphics[width=6cm]{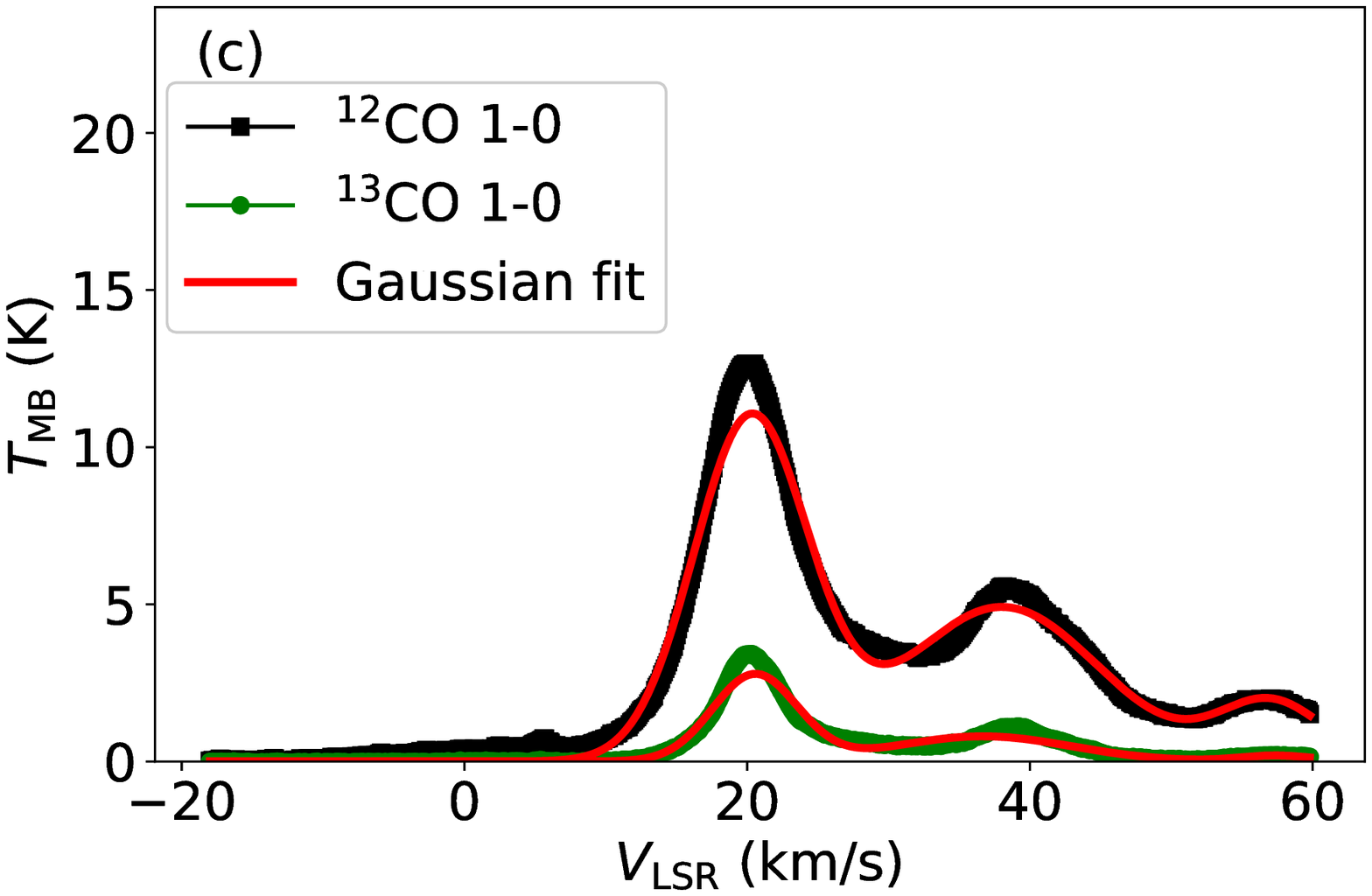} 
\hspace{-0.2cm}

\end{array}$
\vskip -0.3cm
\caption{The $^{12}$CO~($J=1-0$) (blue) and $^{13}$CO~($J=1-0$)  (magenta) spectra averaged over the {\bf (a)} M17-H{\scriptsize II}, {\bf (b)} M17-IRDC, and {\bf (c)} entire M17 regions. Red curves are three-Gaussian component models best fitting the data, and the fitted parameters are listed in Table\,\ref{tab:Gaufit}.}
\label{fig:cospec}
\end{figure*}

\subsection{HCO$^+$ and HCN observations with TRAO 14m telescope}

HCO$^+$ ($J=1-0, 89188.526~{\rm MHz}$) and HCN ($J=1-0, 88631.6023$ MHz) lines were observed simultaneously with the TRAO 14m telescope. The telescope was equipped with the SEQUOIA receiver with 16 pixels in $4\times4$ array. The 2nd IF modules with the narrowband and the eight channels with 4 FFT spectrometers allow observing 2 frequencies simultaneously within the 85--100 or 100--115 GHz frequency ranges for all 16 pixels. We carried out the M17 observations between December 2016 and December 2017. Observations were done in the (On-The-Fly) OTF mode, and the native velocity resolution is about 0.05 km s$^{-1}$ (15 kHz) per channel, and their full spectral bandwidth is 62.5 MHz with 4096 channels. The telescope HPBW beam size is $\sim 50\arcsec$ at 100 GHz and the main-beam efficiency $\eta_{\rm MB}$ is $\sim46\%$ at 89~GHz. The system temperature was in the range 150 -- 300\,K. The cube was regridded to a $7.5\arcsec\times7.5\arcsec$ grid.

\section{Results}
\label{sect:results}

\subsection{Multiple cloud ensemble along the line of sight}

The $^{12}$CO ($J=1-0$) and $^{13}$CO~($J=1-0$) intensity maps integrated over the entire velocity range from -20 to 60\,\kms\, in Figure\,\ref{fig:m17_intall} show the global molecular gas distribution of the M17 complex. The maps cover an area of $1.72\degr\times0.53\degr$ from $13.67\degr$ to $15.39\degr$ in Galactic longitude and from $-0.80\degr$ to $-0.27\degr$ in Galactic latitude. The CO emission around the H{\scriptsize II} region encompassing the NGC~6618 cluster stands out as the brightest subregion in the map. Also, the emission from the M17-IRDC region is  notable in both $^{12}$CO and $^{13}$CO. 

The integrated spectra of $^{12}$CO and $^{13}$CO towards the M17 complex show three major peaks over the complete velocity ranging from -20 to 60\kms\, (see Figure\,\ref{fig:cospec}c). 
The averaged spectra are fitted with three velocity components which show three main components peaking at $\sim20$\,\kms, $\sim38$\kms, and $\sim$57\,\kms; velocity dispersion $\sigma=\frac{FWHM}{2\sqrt{2ln2}}$ of $\sim$4.0\,\kms, 6.9\,\kms, and 3.7\,\kms, respectively. The fitted parameters are summarized in {Table\,\ref{tab:Gaufit}.}  It seems that the main component centers around $\sim $20 \kms\, and the other two components center around $\sim$ 38 \kms\, and $\sim$ 57 \kms.  This becomes more obvious when comparing the $^{12}$CO and $^{13}$CO spectra averaged over the entire M17 complex with those of the individual regions M17-H{\scriptsize II} and M17-IRDC  (Figure\,\ref{fig:cospec}a-b). Each spectrum in these three panels show distinct peaks at different velocities, but the dominant peak is around $V_{\rm LSR} \sim 20\,\kms$, which is the main velocity peak of both M17-H{\scriptsize II} and M17-IRDC regions \citep{elmegreen79}. 
In the average spectrum, we can find at least four groups of molecular clouds in the LSR velocity ranges
$<10$ km s$^{-1}$,
$10-30$ km s$^{-1}$,
$30-50$ km s$^{-1}$,
and
$>50$ km s$^{-1}$.

\begin{figure}
 \begin{center}
 \includegraphics[width=8.7cm]{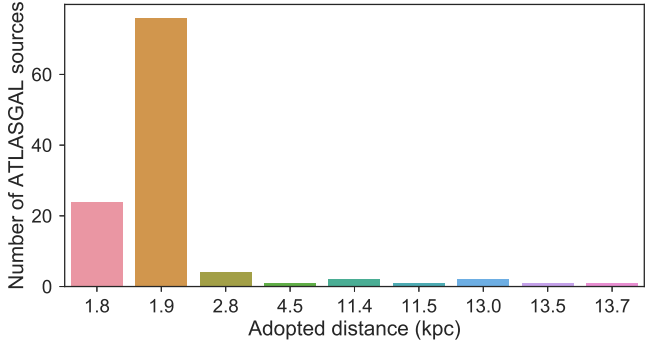}
 \end{center}
\caption{Histogram of distances of ATLASGAL sources extracted from \citet{urquhart18}.}
\label{fig:dist_hist}
\end{figure}

 We further examine the distances of dense clumps in M17 detected with ATLASGAL survey \citep{schuller10,csengeri17}.
The distances were measured by \cite{wienen15} and \cite{urquhart18} using the kinematic method and cross-correlating with maser parallax measurements if available
to resolve the distance ambiguity issues. Approximately 90\% of sources have distances of
1.8 or 1.9\,kpc (Figure\,\ref{fig:dist_hist}), which implies that most clumps in M17 region are located
around $1.9\pm0.1$\,kpc and the region migh have a depth of $\sim0.1-0.3$ kpc.
The distances of dense clumps are close to our assumption
of 2 kpc. The $^{12}$CO and $^{13}$CO position-velocity diagrams show that the emission lines smoothly change from M17-H{\scriptsize II} and M17-IRDC in terms of the radial velocity, line width, and intensity, indicating that the two subregions are physically connected \citep{shimoikura19a}. Thus, the assumption that the two subregions are located at the same distance $\sim$2 kpc is justifiable. 

\begin{table*}[htbp]
\caption{Gaussian parameters best fitting the averaged spectra in Figure\,\ref{fig:cospec}.}
\centering
\begin{center}
\begin{tabular}{ccccccc}
\hline
\hline 

Line & Component & Parameters &  M17 Entire & M17-H{\scriptsize II} & M17-IRDC\\
\hline
\multirow{9}{*}{$^{12}$CO ($J=1-0$)} & \multirow{3}{*}{Gaussian 1} & T [K]		 &10.9   & 9.1    & 10.5 \\
							        &             & $V_{\rm LSR}$  [km s$^{-1}$]   &  20.3 &20.0   & 20.1  \\
								&             & $\sigma_{v}$ [km s$^{-1}$]       & 4.1    &2.3    & 3.7   \\
								\hline
								& \multirow{3}{*}{Gaussian 2} & T [K]	          & 4.9    &4.1    & 5.9  \\
								&             & $V_{\rm LSR}$  [km s$^{-1}$]   & 38.0  &31.3  & 36.6  \\					
								&             & $\sigma_{v}$ [km s$^{-1}$]       & 6.9    &11.7   & 7.4  \\
								\hline
								& \multirow{3}{*}{Gaussian 3} & T [K]	         & 1.9       &1.8    & 2.1 \\
								&             & $V_{\rm LSR}$  [km s$^{-1}$]   & 57.0     &57.0  & 59.3  \\					
								&             & $\sigma_{v}$ [km s$^{-1}$]       & 3.7     &2.4    & 4.7\\		
								\hline		
								\hline						
\multirow{9}{*}{$^{13}$CO ($J=1-0$)} & \multirow{3}{*}{Gaussian 1} & T [K]		& 2.8  	&2.1    &3.5 \\
								&             & $V_{\rm LSR}$  [km s$^{-1}$]   & 20.6	&19.8  &20.2 \\
								&             & $\sigma_{v}$ [km s$^{-1}$]       & 3.0      	& 2.0   & 2.4  \\
								\hline
								& \multirow{3}{*}{Gaussian 2} & T [K]		& 0.8    	& 0.8   &0.9 \\
								&             & $V_{\rm LSR}$  [km s$^{-1}$]   & 36.7	&30.5  &34.1  \\					
								&             & $\sigma_{v}$ [km s$^{-1}$]       & 6.3  	&8.3   &6.9  \\
								\hline
								& \multirow{3}{*}{Gaussian 3} & T [K] 	& 0.2 	&1.6    &0.2\\
								&             & $V_{\rm LSR}$  [km s$^{-1}$]   &57.4 	&68.1    & 57.6  \\					
								&             & $\sigma_{v}$ [km s$^{-1}$]       &2.6 	&1.8    &2.8\\		
								\hline						
\hline

\end{tabular}
\end{center}

\label{tab:Gaufit}
\end{table*}

In Appendix A (Figures \ref{fig:m17_12co_intall} and \ref{fig:m17_13co_intall}), we show the $^{12}$CO and $^{13}$CO~($J=1-0$) intensity maps, respectively, integrated over the velocity ranges from
$-20$ to 10 km s$^{-1}$,
10 to 30 km s$^{-1}$,
30 to 50 km s$^{-1}$,
and
50 to 60 km s$^{-1}$.
The bulk of the M17 emission is seen in the velocity range 10--30\,\kms\, in both $^{12}$CO and $^{13}$CO~($J=1-0$) lines. M17-H{\scriptsize II} region
is especially bright
in the $^{12}$CO~($J=1-0$) maps
and the M17-IRDC region is more prominent in $^{13}$CO~($J=1-0$).
The emission in the range 30--50\,\kms\, is stronger toward the Galactic equator and distributed over a larger region on the plane of the sky.  {\nlq BeSSeL parallax-Based Distance Calculator confirmed that the main velocity component 10--30\,\kms\, is more likely to be in the Sagittaurius arm whereas the 30--50\,\kms\, is more likely to be in between the Scutum arm and Norma arms \citep{reid16,reid19}. Therefore, while the main component is likely at a distance of $\sim2$\,kpc, the 30--50\,\kms\, is in between 3 and 4\,kpc.} 
The emission in the range $>50$\,\kms\, is more scattered and does not appear to correlate with the main bulk emission of M17. The
$^{13}$CO integrated intensity map
in the velocity range 10--30\,\kms\, also coincides with the sub-mm emission observed 
with ATLASGAL (see the three-color image of Figure \ref{fig:m17IR}). 
Subsequently, we consider only the emission around  10--30\,\kms, as a part of M17, and used it to derive physical quantities.

\subsection{Temperature and column density distribution}
\label{sect:CDandT}

Here we derive the excitation temperature, column density, and optical depth of the main cloud component of M17 (10--30 \kms) using our CO data
(see also \citealt{mangum15} for the derivation of these physical quantities).
Based on the
differences
of  $^{12}$CO and $^{13}$CO in the global integrated maps,
we divide these cubes using different masked regions.
Mask 1 is set to the region where both of the
$^{12}$CO and $^{13}$CO integrated intensities
are more than three times as high as the {\nlq noise
levels of 0.22 K \kms\, in the $^{12}$CO integrated map and 0.089 K \kms\, in the $^{13}$CO integrated maps.}
Mask 2 is set to the region where only the
$^{12}$CO emission is detected {\nlq above 3$\sigma$ = 0.27 K \kms}.
Mask 3 is set to the region where
neither 
$^{12}$CO nor $^{13}$CO is detected above 3$\sigma$ levels,
which can be regarded as an emission-free
region. 

We assume that $^{12}$CO and $^{13}$CO having the same excitation temperature $T_{\rm ex}$, and that it can be calculated from the maximum main-beam brightness temperature of $^{12}$CO, 
$T_{\rm max}({\rm ^{12}CO})$, in the masked region 1 as
\begin{equation}
T_{\rm ex} = \frac{5.5\, {\K}}{{\rm ln}\left( 1 + 5.5\,{K}/\left(T_{\rm max}({\rm ^{12}CO}) + 0.82\K  \right)\right)} \,.
\end{equation}
The optical depth $\tau({\rm ^{13}CO})$ as a function of velocity in the masked region 1 can be derived from the main-beam brightness temperature $T_{\rm MB}({\rm ^{13}CO})$ as

\begin{equation}
\tau({\rm ^{13}CO})(v) = - \ln\left(1 - \frac{T_{\rm MB}({\rm ^{13}CO})} {f\left(J(T_{\rm ex}) - J(2.7\,K\right))} \right)\,.
\label{eq:tau13}
\end{equation}
where the filling factor $f=1$ is assumed as the extended nature of the
emission and $J(T_{\rm ex})=\frac{h\nu}{k\left({\rm exp}\left(h\nu/kT_{\rm ex}\right) -1 \right)}
=\frac{5.3}{\left( {\rm exp}\left( 5.3/T{\rm ex} \right)-1\right)}$ for
the $^{13}$CO ($J=1-0$) emission. $h, k, \nu$ are Planck constant,
Boltzmann constant, and transitional frequency.  
Subsequently, the $^{13}$CO column density can be derived as

 \begin{eqnarray}
\label{eq:lsio}
N({\rm ^{13}CO}) & = & \frac{3h}{8\pi^{3} S\mu^{2}} \frac{Q_{\rm rot}}{g_{\rm u}}   \frac{{\rm exp}{(5.3/T_{\rm ex})}}{{\rm exp}{(5.3/T_{\rm ex}-1)}} \int\tau(v)
 dv \,,
\end{eqnarray}
where
$\mu=0.112$ D,
$S=\frac{J_{\rm u}}{2J_{\rm u}+1}$ with $J_{\rm u}=1$, and
$g_{\rm u}=2J_{\rm u}+1=3$.
The rotational partition function is approximated as
$Q_{\rm rot}=\frac{kT_{\rm ex}}{hB}+\frac{1}{3}$ with
$B=55101.011$ MHz with the assumption that all the levels have the same $T_{\rm ex}$ and that $T_{\rm ex}$ is much higher than 5.3~K. This assumption might be wrong but is conventional, which is commonly called the LTE approximation.
We convert the $^{13}$CO column density to H$_{2}$ column density
assuming a
$^{13}$CO fractional abundance of $2\times 10^{-6}$ \citep{dickman78}. We use the updated conversion ratio $N_{\rm H_2}/N_{\rm CO} = 6000$ from \cite{lacy17}, which yielded a $^{13}$CO fractional abundance of $2.5\times 10^{-6}$ using the ratio  $^{12}$C/ $^{13}$C = 60 specifically calculated for M17 \citep{henkel82} and agreed with the average Galactic value \citep{langer90}.

For mask 2 region, we use the $^{12}$CO ($J=1-0$) emission as a proxy to estimate the H$_{2}$ column density assuming that the emission is optically thin as $N_{\rm H_{2}}=X W({\rm CO})$ where $X$ is the conversion factor and $W${\rm (CO}) is the $^{12}$CO ($J=1-0$) integrated intensity. The $X$ factor $X=2\times10^{20}~{\rm cm}^{-2}$ ${\rm K}^{-1}\,(\rm \kms)^{-1}$ is used as recommended by \citet{bolatto13}, which was established after an exhaustive investigation of all possible measurements. Note that we do not estimate $T_{\rm ex}$ for this masked region.

\begin{figure*}[!htbp]
$\begin{array}{ccc}
\hskip -0.3cm
\includegraphics[width=6.1cm]{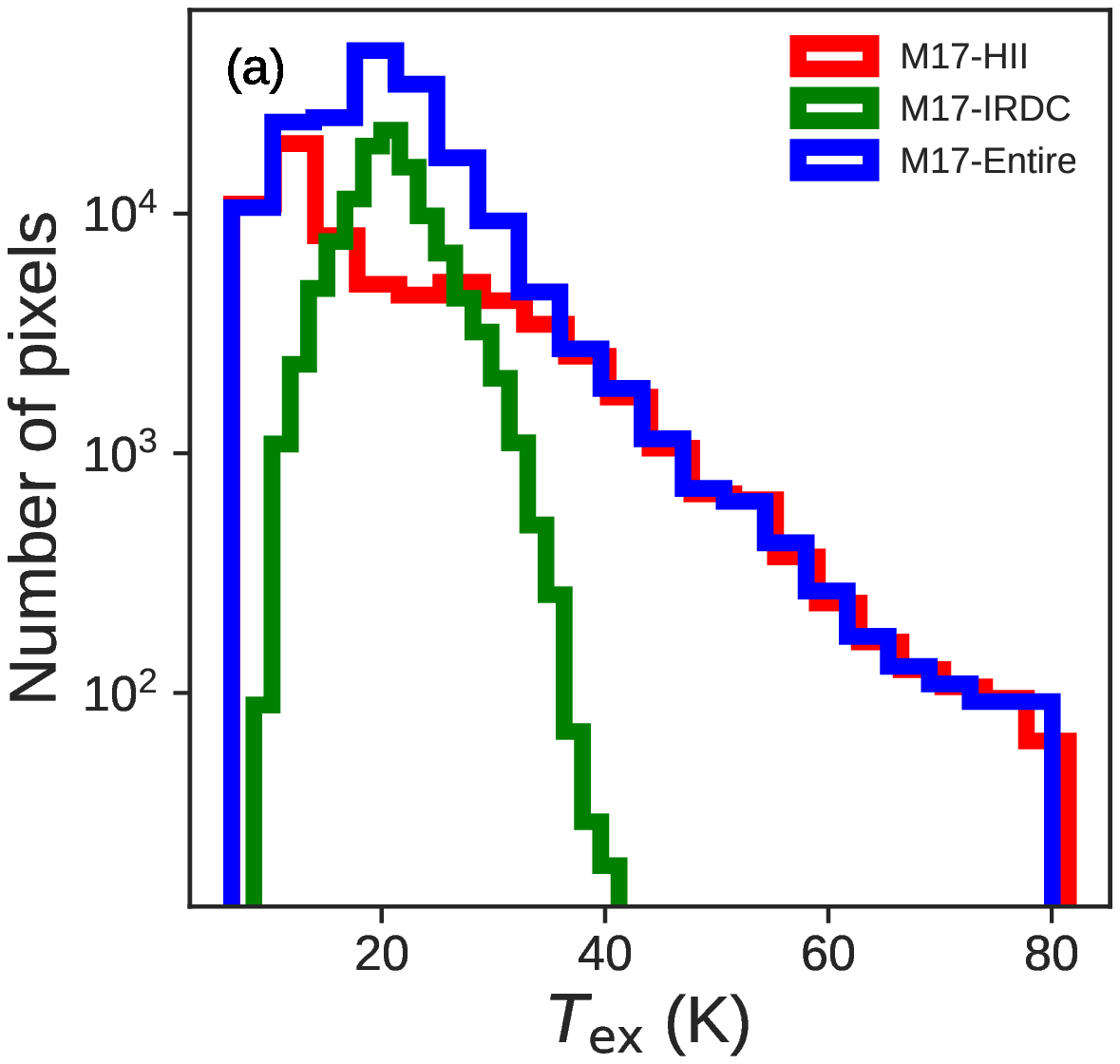} &
\hskip -0.3cm
\includegraphics[width=6.1cm]{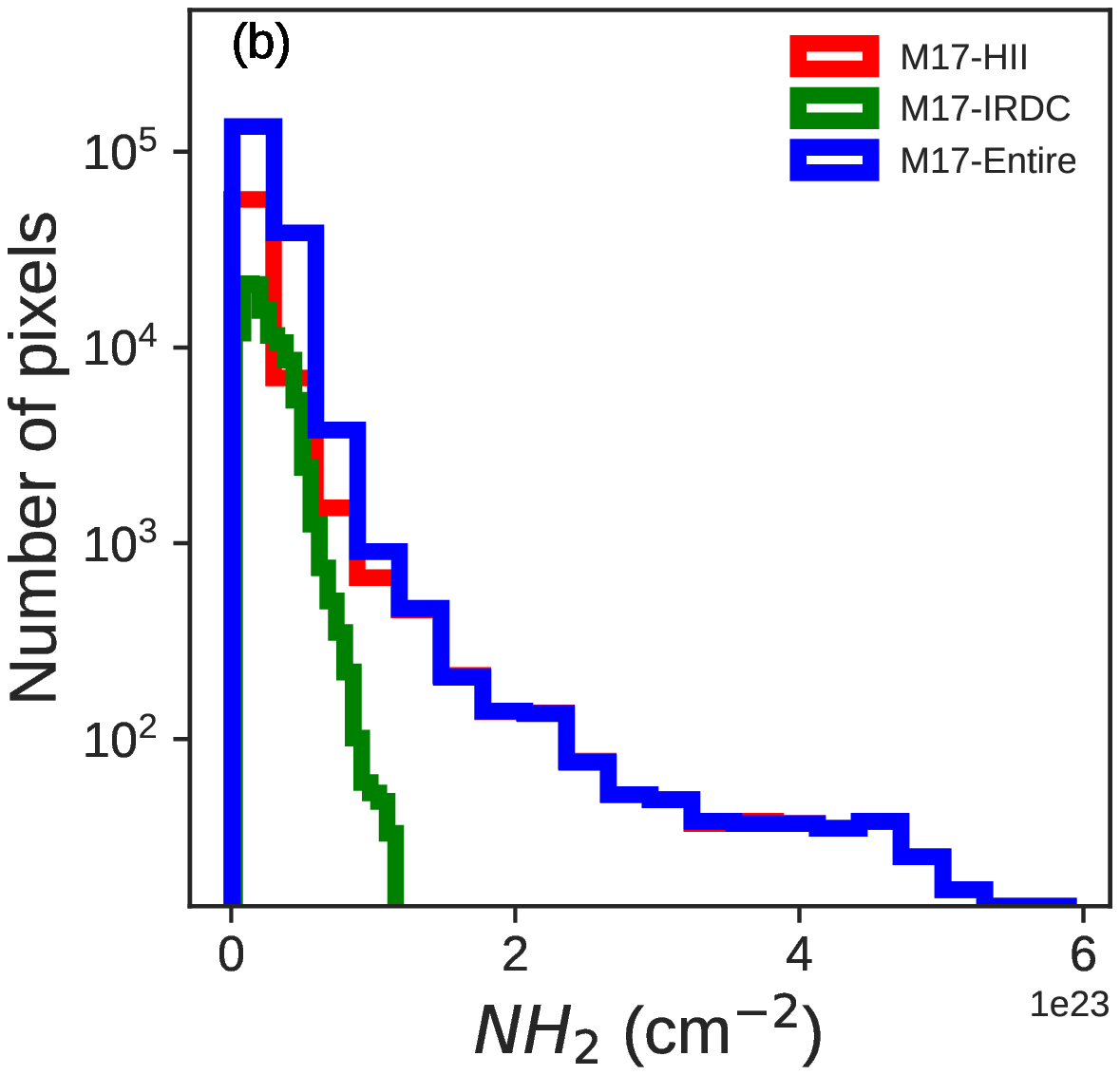}&
\hskip -0.3cm
\includegraphics[width=6.1cm]{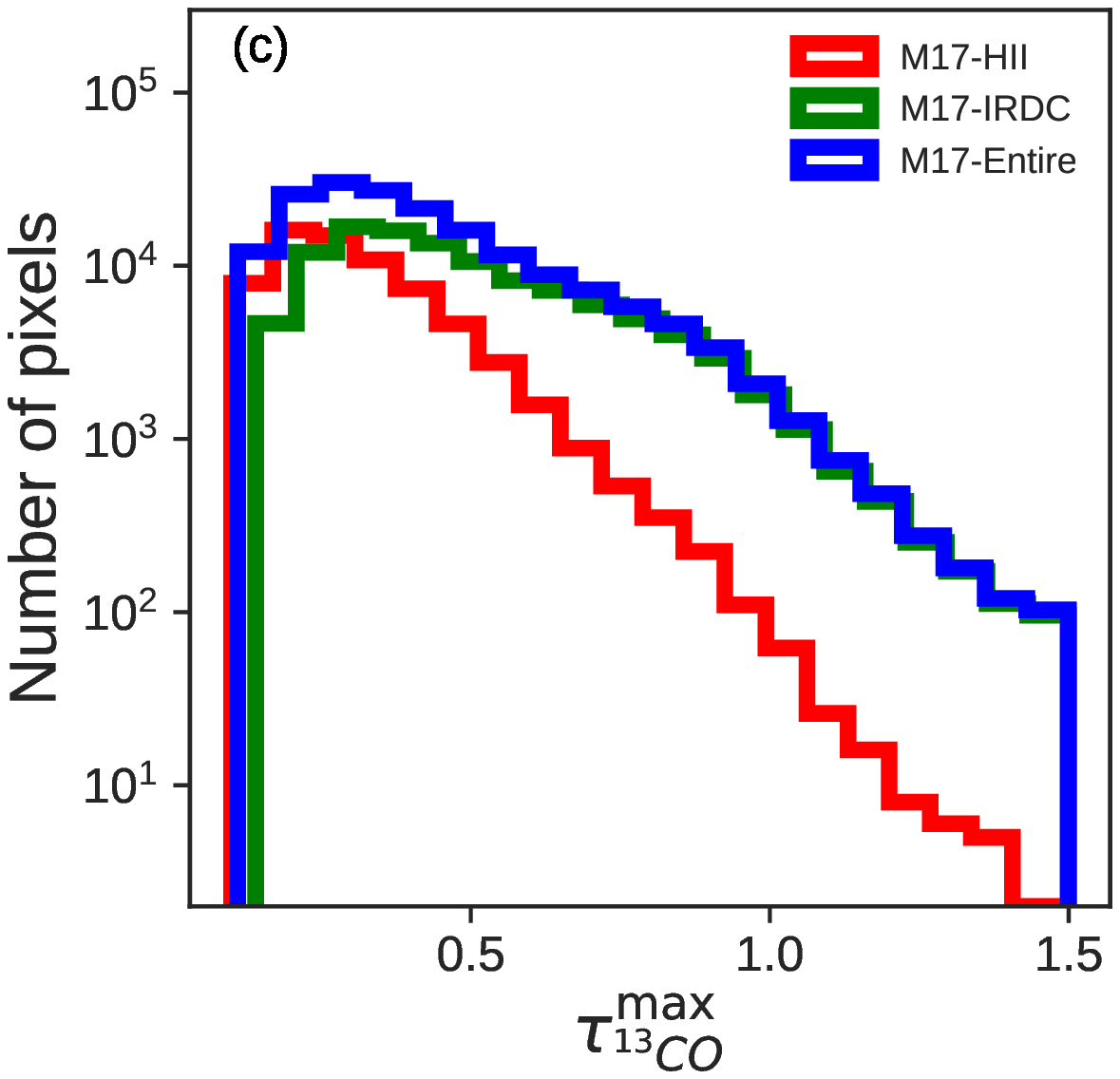}\\

\\
\end{array}$
\vskip -0.3cm
\caption{The histograms of the {\bf (a)} CO excitation temperature, {\bf (b)} the H$_2$ column density, and {\bf (c)} the peak optical depth of the M17 region.
The blue, orange, and green histograms are toward   the entire M17, M17-H{\scriptsize II}, and M17-IRDC, respectively.}
\label{fig:hist}
\end{figure*}

\begin{figure*}[htbp]
 \begin{center}
  \vskip 0cm
 \includegraphics[width=17cm]{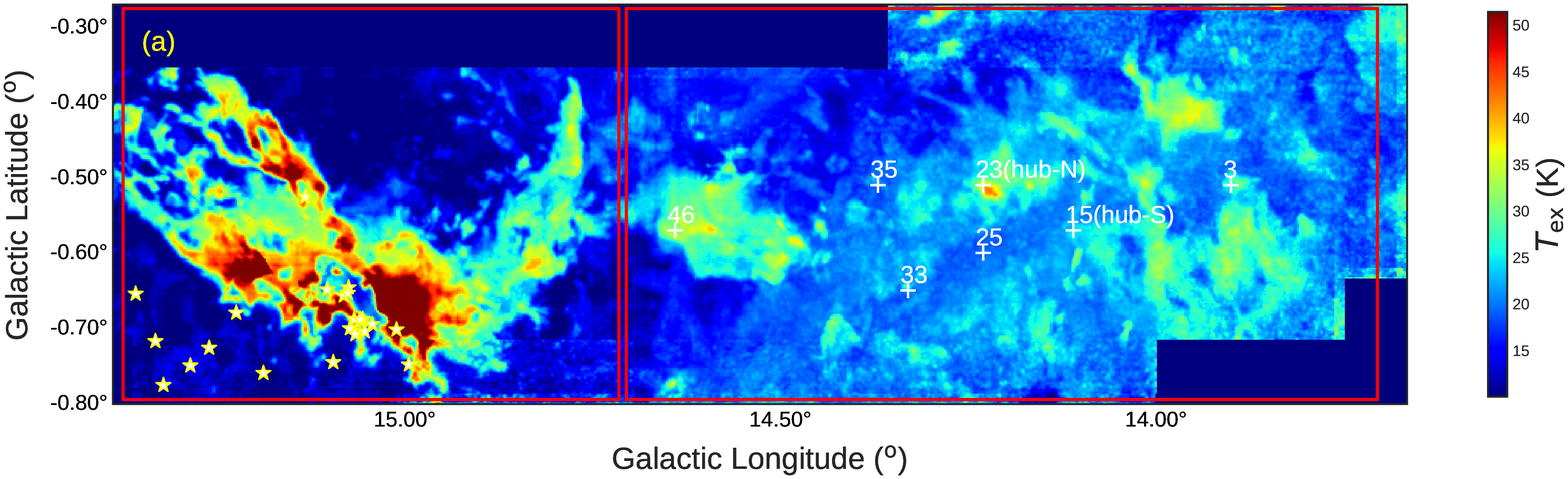}\\
  \includegraphics[width=17cm]{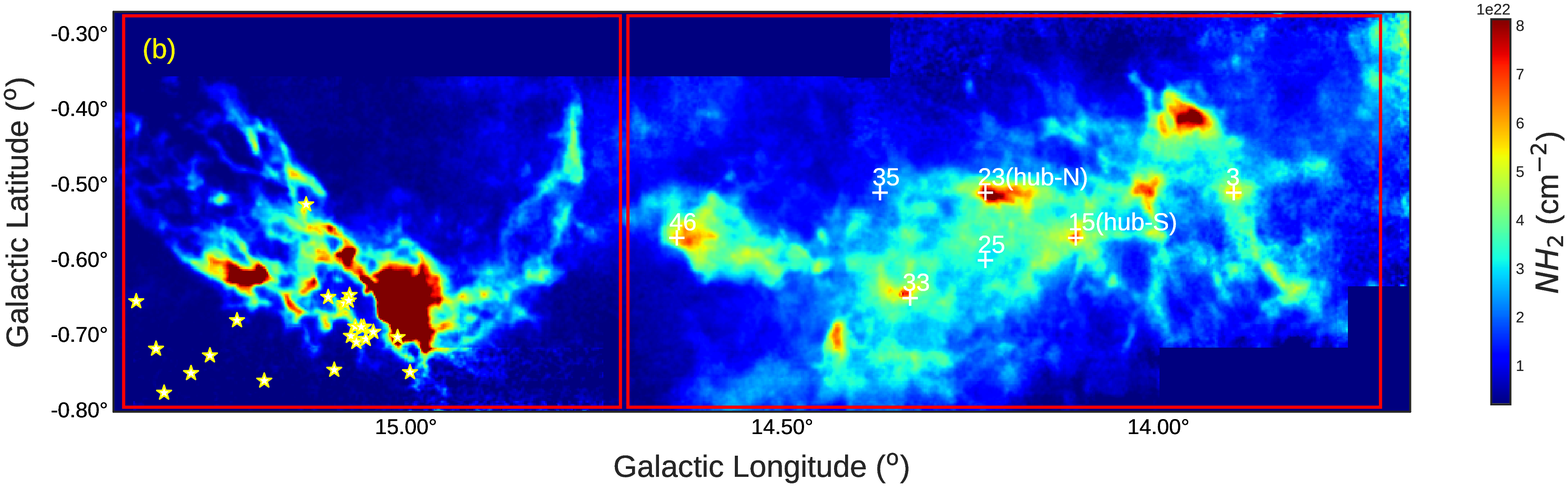}\\
   
 \includegraphics[width=17cm]{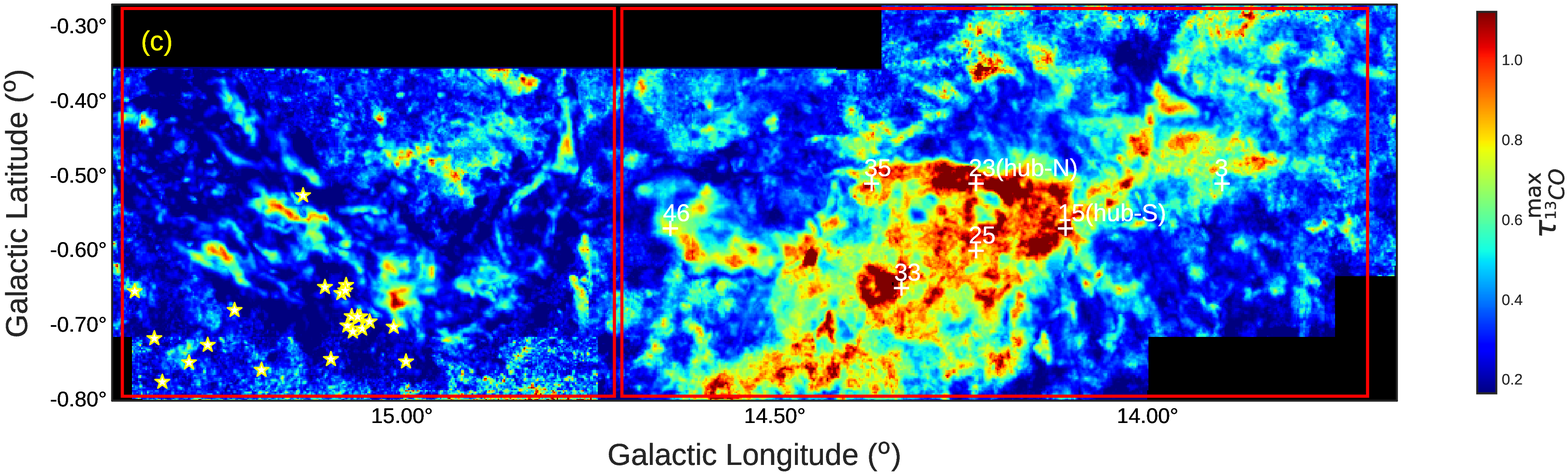}\\

  \end{center}
\caption{The maps of the {\bf ({a})} CO excitation temperature, {\bf ({b})} H$_2$ column density, and {\bf ({c})} peak optical depth of the M17 complex. We use only data within the main velocity range (10--30\,\kms) of M17. The red rectangles outline the two prominent star-forming regions: M17-H{\scriptsize II} (left) and M17-IRDC (right). {The star symbol pinpoints the OB star cluster responsible for the giant H{\scriptsize II} region. The cross symbols pinpoint the locations of the massive cores (M $>$ 500\,\msun) reported by \cite{shimoikura19a}.}}
\label{fig:T_CDmaps}
\end{figure*}

Excluding pixels where excitation temperature is less than 2.7\,K (14\% number of pixels) caused by low rms in the integrated maps and using only non-zero data within the percentile range 0.5\%--99.99\%, we obtain a temperature range of 8 -- 81\,K, and an  $N_{\rm H_{2}}$  range of $3\times10^{20}$-- $4.6\times10^{23}$\,cm$^{-2}$ for the entire M17 complex (Figures \ref{fig:hist} and \,\ref{fig:T_CDmaps}). 
The median CO excitation temperature is ~$\sim 20$\,K, and median column density is $1.3\times10^{22}$\,cm$^{-2}$ for the entire M17 (Table\,\ref{tab:hist}).

\begin{table*}[htbp]
\caption{Properties of the sub-region M17-H{\scriptsize II}, M17-IRDC and the M17 complex. $T_{\rm ex}$, $N_{\rm H_{2}}$, and  $\Sigma_{\rm gas}$ are measured per pixel.}
\begin{center}
\begin{tabular}{cccccccccccccc}
\hline
\hline
Region & ($T_{\rm ex}^{\rm min}, T_{\rm ex}^{\rm median}, T_{\rm ex}^{\rm max}$) & ($N_{\rm H_{2}}^{\rm min}, N_{\rm H_{2}}^{\rm median}, N_{\rm H_{2}}^{\rm max}$) & ($\Sigma^{\rm min}_{\rm gas}, \Sigma^{\rm median}_{\rm gas}, \Sigma^{\rm max}_{\rm gas}$) &$A_{\rm tot}$  & $M_{\rm tot}$  & $n^{\rm mean}_{\rm H_{2}}$ \\
& (K) & $(10^{21}$\, cm$^{-2})$ & (\msun pc$^{-2})$ & (pc$^{2}$) & ($\msun$) & (cm$^{-3}$)\\
\hline
M17-H{\scriptsize II} & (8, 15, 81)      & (0.3, 6.3, 458) & (6, 134, 9831) & 364 & $1.43\times10^{5}$ & 166\\
M17-IRDC & (11, 21, 41) & (1.9, 17, 93) &(40, 355, 1995) & 593 &$3.06\times10^{5}$ & 285\\
M17 & (8, 20, 81)           & (0.3, 13, 458) & (6,281,9831) & 957 & $4.49\times10^{5}$ & 201\\
\hline
\hline
\end{tabular}
\end{center}
\label{tab:hist}
\end{table*}

{\nlq There are strong difference between M17-H{\scriptsize II} and M17-IRDC. First, the peak optical depth $\tau_{\rm ^{13}CO}^{\rm max}$ is remarkably different in two regions as seen in Figure~\ref{fig:hist} and \ref{fig:T_CDmaps}.  $\tau_{\rm ^{13}CO}^{\rm max}$ in M17-H{\scriptsize II} has a mean value of 0.31 and a standard deviation of 0.15 while those of M17-IRDC are 0.5 and 0.24, respectively. Some regions in M17-IRDC have $\tau_{\rm ^{13}CO}^{\rm max}$ values that are even higher than 1. 
Second, M17-H{\scriptsize II} has much higher temperature than M17-IRDC. The maximum temperature in M17-H{\scriptsize II} region reaches 81\,K, especially around the NGC~6618 cluster, whereas the maximum temperature in M17-IRDC is only 41\,K} (Figure \ref{fig:T_CDmaps}a). Third, beside being colder, M17-IRDC also has lower peak column density compared to M17-H{\scriptsize II} (Figure\,\ref{fig:T_CDmaps}b). We caution that the column density is probably over-estimated because the partition function is overestimated due to the fact that $T_{ex}$ is not much greater than 5.5 K.

\subsection{Dense gas mass function and mass properties}
\label{sect:cummilative}
\begin{figure*}[!htbp]
 $\begin{array}{c}
\includegraphics[width=16cm]{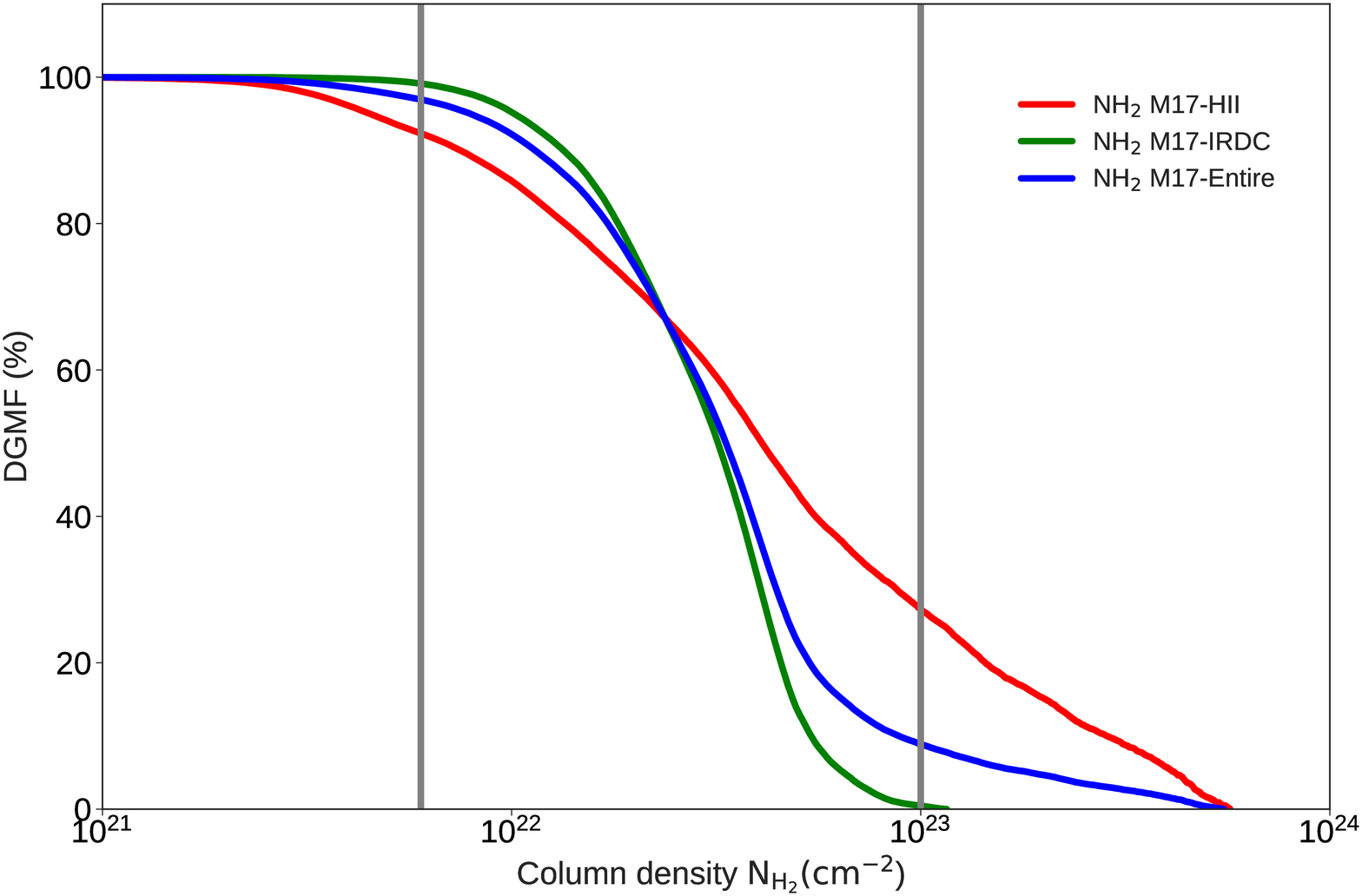} \\
\end{array}$
\caption{The normalized cumulative column density distributions
of the entire M17, M17-H{\scriptsize II},  
and M17-IRDC regions.
The vertical lines mark the locations of the column densities of  $7\times10^{21}$cm$^{-2}$ and $1\times10^{23}$cm$^{-2}$. }
\label{fig:DGMF}
\end{figure*}

Figure \ref{fig:T_CDmaps}b clearly shows that M17-H{\scriptsize II}
has higher column density materials than M17-IRDC.
The dense gas mass function (DGMF) is useful to see how the dense gas is concentrated in specific density ranges.
Dividing the column density into 120 column density bins ranging from $10^{21}$\,\cmd\, to $10^{24}$\,\cmd\, we create the DGFMs as the normalized Cumulative Mass Distribution (CMD) as:
\begin{equation}
{\rm DGMF(>N_{\rm H_2})} =\frac{\int_{N_{\rm H_2}}^{N_{\rm H_2}^{\rm max}} \mu m_{\rm H} A({N_{\rm H_2}})  dN_{\rm H_2}}{M_{\rm tot}} \ ,
\label{eq:CMD}
\end{equation}
where $N_{\rm H_2}$ is the column density from which the mass is accumulated, $\mu=2.8$ is the mean molecular weight, $m_{\rm H}$ is the hydrogen atomic mass, $A$ is the integrated area, and $N_{\rm H_2}^{\rm max}$ is the maximum column density observed in the region. When $N_{\rm H_2}$ reaches the noise level $N_{\rm H_2}^{\rm min}$, the obtained mass is the total mass of the cloud $M_{\rm tot}$ in both masks 1 and 2. $N_{\rm H_2}^{\rm min}$ is the column density corresponding to the 3$\sigma$ (Table\,\ref{tab:hist}).  
DGMFs for M17, M17-H{\scriptsize II}, and M17-IRDC are plotted in a log-log scale in Figures~\ref{fig:DGMF}. They have flat profiles below the column density $\sim 7\times10^{21}$~\,\cmd\, and then a quick drop to a power-law tail at higher column density. The column density $\sim 7\times10^{21}$~\,\cmd\, is argued to be the threshold for star formation \citep{andre10,arzoumanian10,lada10,heiderman10,evans14}. The slope of the power-law tail is shallower in M17-H{\scriptsize II} region than in M17-IRDC region. DGMFs converge to unity toward lower column density but diverge at higher column density. These profiles are similar to CMDs and DGMFs of other regions (\citealt{lada10}, \citealt{kainulainen13}, and \citealt{rivera-ingraham15}).

{\nlq We calculate the very dense gas mass fraction, the ratio of mass that has column density higher than $1\times10^{23}\,\cmd$ or  1g/cm$^2$, a column density level suggested as massive star formation threshold \citep{krumholz08} as:
\begin{equation}
f_{\rm very dense} = \frac{M_{\rm N_{H2}>1\times10^{23}\,\cmd}}{M_{\rm tot}}
\end{equation}
In total, $f_{\rm very dense}$ is about 8.86\% in the entire M17. Individually, $f_{\rm very dense}$ is 27\% in M17-H{\scriptsize II} and only 0.46\% in M17-IRDC region.} 

On the other hand, the DGMFs for mass with column density above the presumed threshold for star formation $7\times10^{21}~\,\cmd$\, are 96\%, 91\%, and 98.5\% for M17, M17-H{\scriptsize II}, and M17-IRDC, respectively. These two column density levels are marked as vertical lines in Figures~\ref{fig:DGMF}. 

The flat plateau of the DGMF at the low column density part gives us the total masses of M17, 
M17-H{\scriptsize II} and M17-IRDC (Figure\,\ref{fig:DGMF}) which are captured in Table\,\ref{tab:hist}. The total mass of M17 is $4.49\times10^{5}$\,\msun, that of 
M17-H{\scriptsize II} is $1.43\times10^{5}$\,\msun, and that of M17-IRDC is 
$3.06\times10^{5}$\,\msun. The median pixel-by-pixel mass surface density of M17-IRDC (614\,\msun\,pc$^{-2}$) is higher than that of the M17-H{\scriptsize II} region (280\,\msun\,pc$^{-2}$). However, the peak mass surface density is much higher in M17-H{\scriptsize II} (9831\,\msun pc$^{-2}$) than in M17-IRDC (1995\,\msun pc$^{-2}$) (see Table\,\ref{tab:hist}).  The mean column density $N^{\rm mean}_{\rm H_{2}}$ is calculated as  

\begin{equation}
N^{\rm mean}_{\rm H_{2}}=\frac{M_{\rm tot}}{A_{\rm tot}}
\end{equation}
where $A_{\rm tot}$ is the total areas of the clouds {\nlqb calculated as the sum of areas of all interior pixels of the cloud projection on the plane of the sky}. We also calculate the volume density  as $n_{H_2} = \frac{M}{V} = \frac{M}{\frac{4}{3}\pi r^{3}}$ where $r =\sqrt{\frac{A_{\rm tot}}{\pi}}$ and $A_{\rm tot}$ is tabulated in Table\,\ref{tab:hist}. The mean volume density is higher in M17-IRDC (285\,cm$^{-3}$) than in M17-H{\scriptsize II} (166\,cm$^{-3}$). For the two column density levels discussed above, the corresponding average volume densities of M17-H{\scriptsize II}, M17-IRDC, and entire M17, respectively, are $1.0\times10^{5}$, $2.1\times10^{5}$, and $9.5\times10^{4}$ cm$^{-3}$ for column density above $1\times10^{23}\,\cmd$, and $3.5\times10^{3}$, $1.7\times10^{3} $, $1.5\times10^{3}$ cm$^{-3}$ for column density above $7\times10^{21}\,\cmd$.

In summary, the peak column density is higher in M17-H{\scriptsize II} while the average column density and total mass are higher in M17-IRDC. If very dense gas is considered alone, its average volume density is approximately two times more in M17-H{\scriptsize II} than in M17-IRDC. 

\begin{table*}[htbp]
\caption{Statistics of individual clumps extracted with the Dendrogram technique
in the M17-H{\scriptsize II}, M17-IRDC, and the entire M17 clouds}
\begin{center}
\begin{tabular}{ccccccccccccc}
\hline
\hline
Region & $n_{\rm sources}$  & $n_{\rm sources}( \alpha_{\rm vir} >1)$ & ($R^{min}, R^{median}, R^{max}$)  &  ($M^{min}, M^{median}, M^{max}$)  & ($\alpha_{\rm vir}^{min}, \alpha_{\rm vir}^{median}, \alpha_{\rm vir}^{max}$)  \\
 &   &  & pc  &  $M_\odot$  &   \\
\hline
M17-H{\scriptsize II}  &  26 & 11 & (0.12, 0.2, 0.41) & (3, 60, 1719) & (0.12, 0.86, 8.23) \\
M17-IRDC &  164 & 105 & (0.11, 0.2, 0.54) & (1, 14, 1001) & (0.14, 1.36, 14.23) \\
M17-Entire &  190 & 116 & (0.11, 0.2, 0.54) & (1, 17, 1719) & (0.12, 1.33, 14.23) \\

\hline
\hline
\end{tabular}
\end{center}
\label{tab:Dendrogram}
\end{table*}

\subsection{Individual clump structure extracted with Dendrogram}

To obtain an automatic extraction of  the individual clumps in M17, we use the Dendrogram  \citep{rosolowsky08}\footnote{https://Dendrograms.readthedocs.io/en/stable/}. It is a hierarchical clustering method that builds up clusters in a tree-like structure where each node represents a leaf (structure that has no sub-structure) or branch (structure that has successor structure).  Each node in the cluster tree contains a group of similar data. Clusters at one level join with clusters in the next level up using a degree of similarity. The total number of clusters is not predetermined.

In our extraction, we use  Dendrogram to detect and extract the morphologies of individual clumps using the $^{12}$CO (J = 1-0) data cube in the velocity range of 10.0 to 30.0 km s$^{-1}$. We extract only sources in the regions that have column density above $3\times10^{22}\, \rm cm^{-2}$ as in the masked map (Figure\,\ref{fig:Dendrogram}). This threshold corresponds to 3 times the median column density of the entire map {\nlqb ({Table\,\ref{tab:hist}})}.
Dendrogram requires three input parameters: 
\texttt{min\_value}, \texttt{min\_delta}, and \texttt{min\_npix}.
The first parameter \texttt{min\_value} specifies the minimum value above which Dendrogram
attempt to identify the structures.  The second parameter \texttt{min\_delta}  is the minimum step 
required for a structure identified. The third parameter \texttt{min\_npix} is 
the minimum number of pixels that a structure should contain in order to remain 
an independent structure.
We set the three parameters as follows:  
\texttt{min\_value} $= 5\sigma$, \texttt{min\_delta}=$3 \sigma$, and \texttt{min\_npix} = 100, where
$\sigma$ is the average rms noise level of the $^{12}$CO data.{\nlqb We select \texttt{min\_npix}=100 which is in the middle of 4x4x4=64 and 5x5x5=125 voxels in position-position-velocity space in order to remove artificial structure in the area having high noise after some trials. }The map angular resolution is close to about 4 times the cell size. We keep only leaf structures, which are independent structures in our analysis.
This extraction with Dendrogram results in the identification of
{\nlq 26 individual clumps in M17-H{\scriptsize II} and 164 clumps in M17-IRDC. The properties of these clumps can be found in Table\,\ref{tab:Dendrogram}. 
Here, we assumed all the distances to the structures identified are equal to the representative 
distance of $d=2$ kpc.}

We use the virial parameters as the ratio of the virial mass to the true clump mass,  `$\alpha_{\rm vir}$', as a measure of the gravitational
stability of the clumps. For clumps in gravitational equilibrium, $\alpha_{\rm vir}$ is unity.
Collapsing and  dispersing clumps have $\alpha_{\rm vir}$ of $<1$ and $>1$, respectively.
The virial parameter for a spherical clump with uniform density and temperature  neglecting external pressure and magnetic field 
can be expressed as
\begin{equation}
\alpha_{\rm vir} =\frac{M_{\rm vir}}{M_{\rm clump}}  =\frac{5\sigma_{\rm 3D}^{2}R}{3GM_{\rm clump}}  \, ,
\label{eq:virial}
\end{equation}
where $G$, $M$ and $R$ are the gravitational constant, clump mass and clump radius, respectively \citep{bertoldi92}.
The clump radius $R$ is defined as the geometric mean of the major and minor semi-axes of the projection onto the position-position plane for a clump identified. 
3D velocity dispersion $\sigma_{\rm 3D}$ is calculated as $\sigma_{\rm 3D} =  \sigma_{\rm 1D}/\sqrt{3}$.
The clump mass is the LTE mass calculated from the $^{12}$CO ($J = 1-0$) integrated intensity of the individual structures.


The masses of the individual clumps extracted with Dendrogram in M17-H{\scriptsize II} appear to be comparable to those in M17-IRDC while the peak column density in M17-H{\scriptsize II} is higher than in M17-IRDC as seen in the histogram of column density (Figure\,\ref{fig:DGMF}). 
The median clump mass is 60\, and 14\,\msun\, in M17-H{\scriptsize II} and M17-IRDC, respectively. 
Both clumps in M17-H{\scriptsize II} and M17-IRDC have a median radius of 0.2~pc.
The virial parameters of clumps in M17-H{\scriptsize II} are smaller than those in M17-IRDC. 64\% of clumps in M17-IRDC has virial parameteres $\alpha_{\rm vir} > 1$ while 42\% of clumps in M17-H{\scriptsize II} has $\alpha_{\rm vir} > 1$. However, the median virial parameters in M17-IRDC is 1.36, higher than the median value 0.86 in M17-H{\scriptsize II}. Thus, the clumps in M17-H{\scriptsize II} are more prone to gravitational contraction, which is consistent with the fact that star formation is much more active in M17-H{\scriptsize II}.

\subsection{Distribution of dense gas traced by HCO$^{+}$ and HCN}

\begin{figure*}[htbp]
 \begin{center}

  \includegraphics[width=17cm]{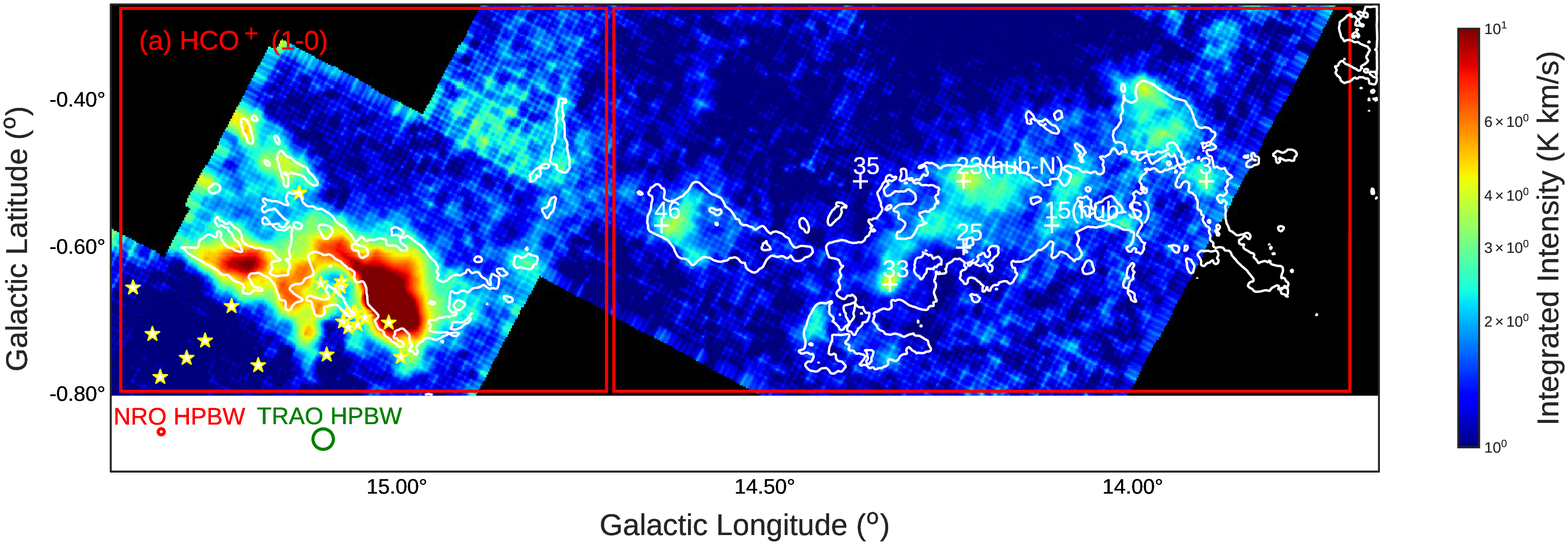}\\
 \includegraphics[width=17cm]{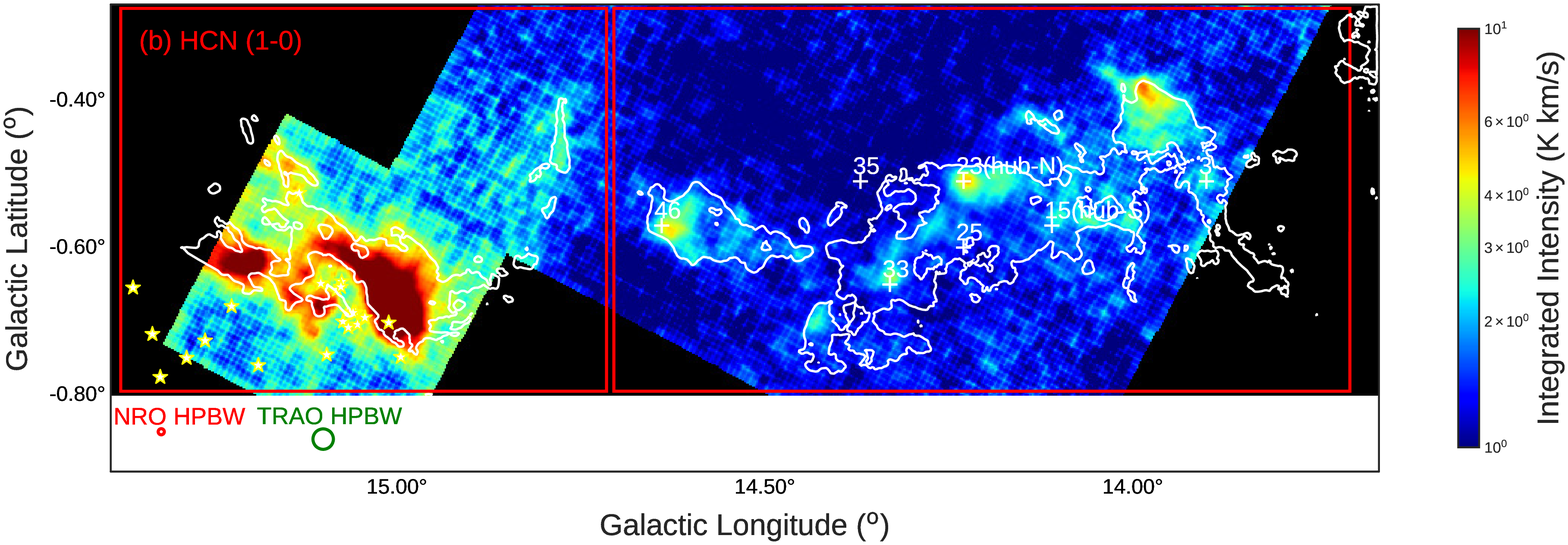}
  \vskip -1cm
 \end{center}
\caption{{\bf (a)} The HCO$^{+}$ ($J=1-0$) and {\bf (b)} HCN ($J=1-0$) integrated intensity maps of M17. The red rectangles outline the two prominent star-forming regions: M17-H{\scriptsize II} {\bf (left)} and M17-IRDC {\bf (right)}. {The star symbols pinpoint the OB star clusters responsible for the giant H{\scriptsize II} region.} The cross symbols pinpoint the locations of the massive cores (M $>$ 500\,\msun) reported by \cite{shimoikura19a}. White contours correspond to the gas column density level of $3\times10^{22}\,\cmd$.}
\label{fig:m17_hcop_hcn}
\end{figure*}

M17-H{\scriptsize II} and M17-IRDC have different high column density gas concentration seen in column density maps created from CO ($J=1-0$) observations as discussed in Section~\ref{sect:CDandT}. We use  HCO$^{+}$ ($J=1-0$) and HCN ($J=1-0$) to further examine the distribution of high density gas in M17, which are thought to be better tracers of dense star-forming gas than CO isotopologues \citep{gao04, wu10, stephens16}. The optically thin critical densities
of HCO$^{+}$ ($J=1-0$) is as high as $6.8\times10^4$ cm$^{-3}$ and that of HCN ($J=1-0$) is as high as $4.7\times 10^{5}$ cm$^{-3}$ at 10 K \citep{shirley15}. However, the effective critical densities at the same temperatures are $9.5\times10^2$ cm$^{-3}$ for HCO$^{+}$ ($J=1-0$) and $8.4\times10^3$ cm$^{-3}$ for HCN$^{+}$ ($J=1-0$). For comparision, the critical density of $^{12}$ CO ($J=1-0$) is less than 10$^{3}$ cm$^{-3}$  \citep{shirley15}. 
Beside tracing the general landscape of dense gas, these tracers are sensitive to the physical processes closely related to star formation such as outflows, infall, photoionization, mechanical energy, and chemistry \citep{sanhueza12,walker-smith14,chira14,fuller05,roberts11,shimajiri17}.

In Figure~\ref{fig:m17_hcop_hcn}, we present the integrated intensity maps of HCN and HCO$^{+}$ ($J=1-0$) transitions from 10 to 30\,$\kms$. The $1\sigma$ levels are $\sim1\,\K \kms$ and $1.2\,\K\kms$ for the HCN and HCO$^{+}$ integrated maps. It clearly shows that most of the emission of these lines is in M17-H{\scriptsize II} region, indicating that M17-H{\scriptsize II} region contains a significant amount of high density gas.  
The HCO$^+$ and HCN emission come mainly from the high column density ($> 3\times10^{22}\,\cmd$ or 600 \msun pc$^{-2}$) part traced in CO except near the star cluster where the column density is lower than $3\times10^{22}\,\cmd$ but a small fraction of HCO$^{+}$ and HCN ($J=1-0$) is still detected. 
In contrast, the emission of HCO$^{+}$ and HCN in M17-IRDC is almost invisible except in a few dense clumps and intersections of the IRDC filamentary network \citep{busquet16}, indicating that there are more dense gas at the intersections of the filaments \citep{ohashi16}. These locations coincide with the bright $^{13}$CO emission positions (see Figure\,\ref{fig:m17_13co_intall}). {\nlq Note that these areas of both HCN and HCO$^{+}$ emission are similar and are the sums of all pixels that have HCN and HCO$^{+}$ emission larger than zero. The total areas of HCN/HCO$^{+}$ of H{\scriptsize II} and M17-IRDC are 23.6\,pc$^{2}$ and 52.6\,pc$^{2}$, respectively. Thus the HCN and HCO$^{+}$ emitting areas are 6.5\% and 8.9\% of the total CO emitting areas for  H{\scriptsize II} and M17-IRDC, respectively.} While the trends are similar, HCN integrated intensity is higher than that of HCO$^{+}$.
The maximum HCO$^{+}$ integrated intensity in M17-IRDC is about {5\,K\,\kms} while that of M17-H{\scriptsize II} goes up to 35\,K\,\kms.

\begin{figure*}[!htbp]
\vskip -0.1cm
\hspace{-0.3cm}
 $\begin{array}{cc}
\includegraphics[width=8.5cm]{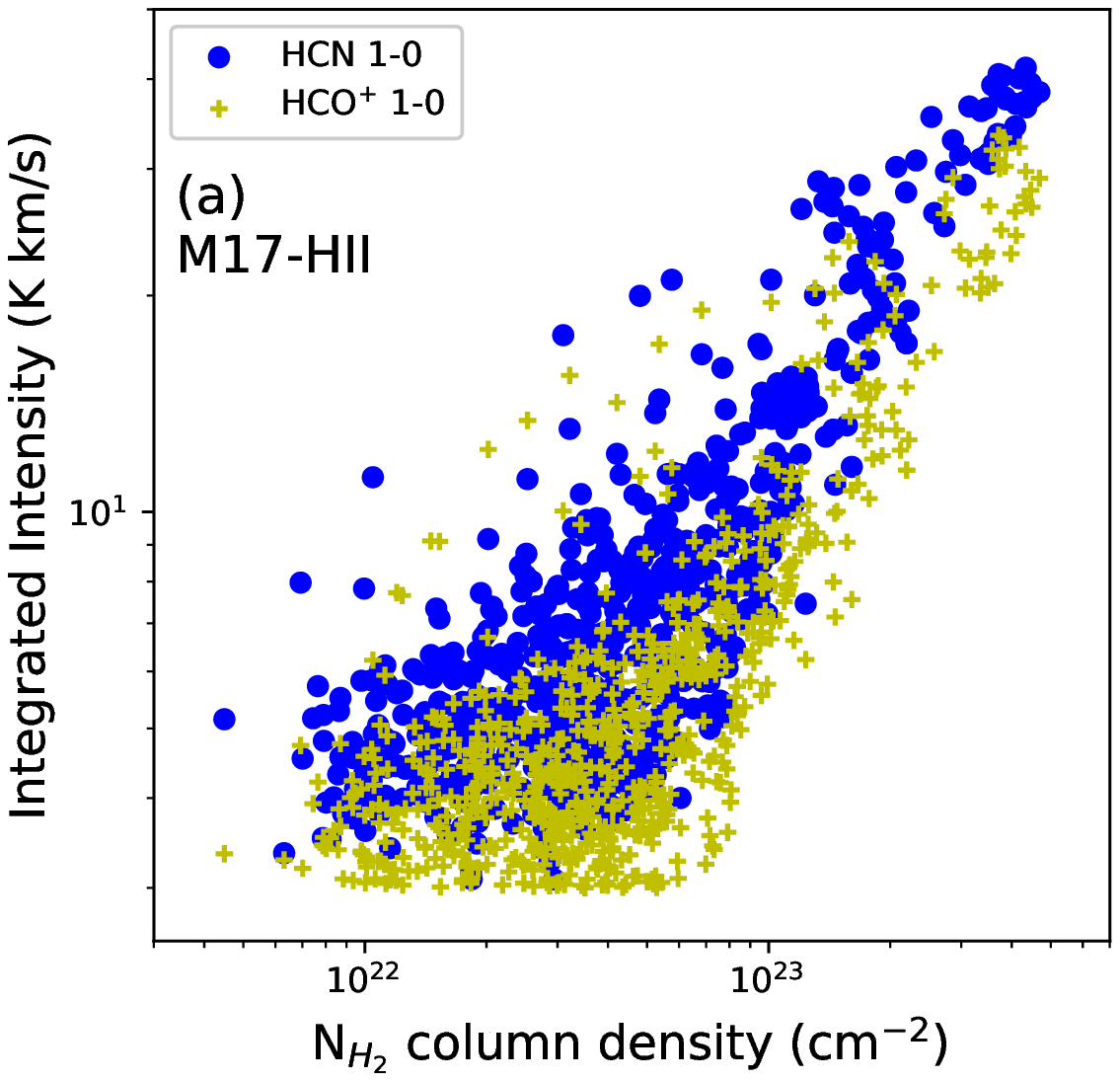} &
\includegraphics[width=9cm]{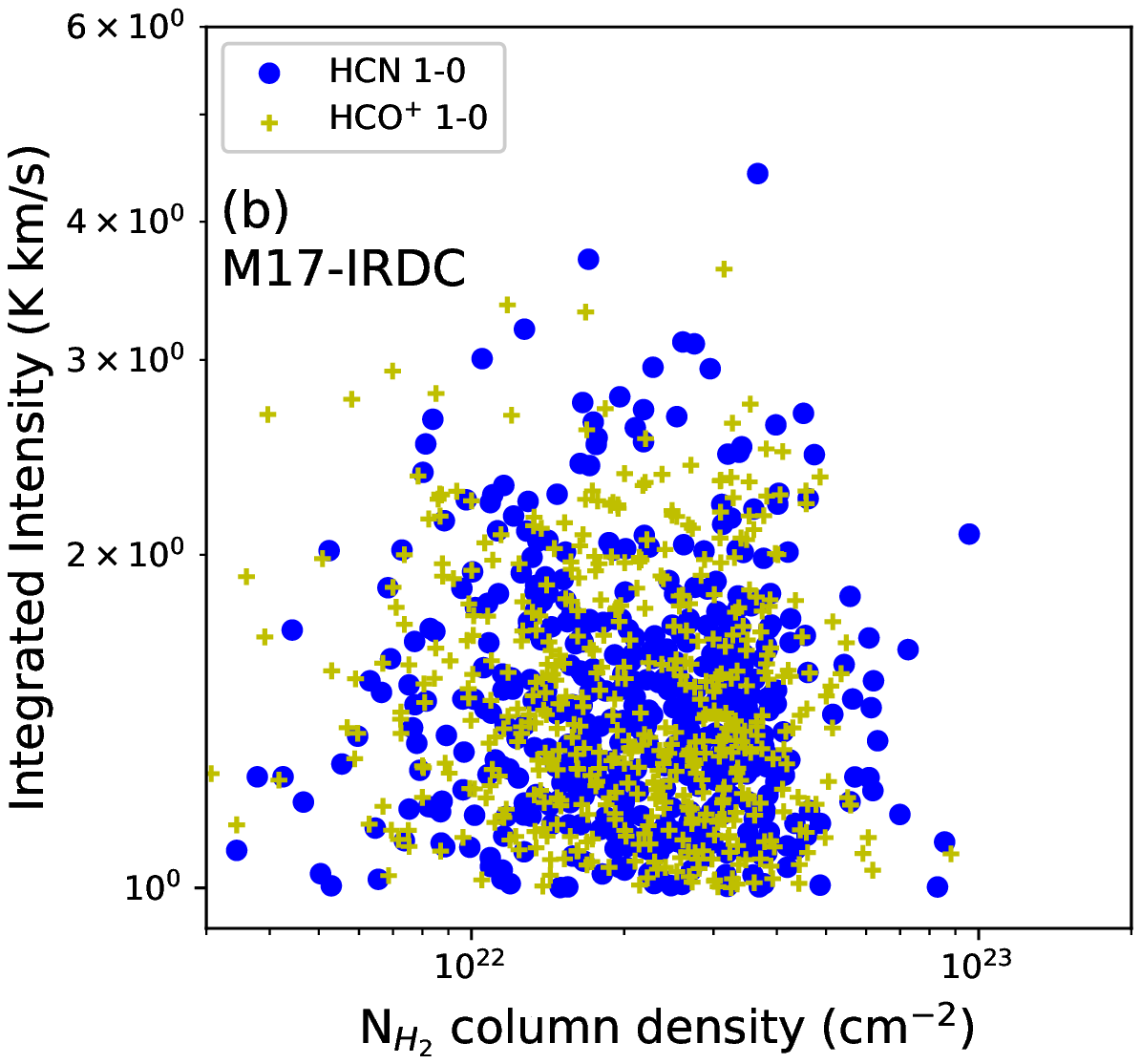} 
 \\
\hspace{-0.3cm}
\\
\end{array}$
\caption{
The scatter plots of pixel-by-pixel HCN and HCO$^+$ ($J= 1-0$) integrated intensity versus gas column density for {\bf (a)} M17-H{\scriptsize II} and {\bf (b)} M17-IRDC regions. We note that the sacle are different in these plots.
}
\label{fig:m17dense}
\end{figure*}

While the integrated intensities of HCN and HCO$^{+}$ depend linearly on the gas column density in M17-H{\scriptsize II} cloud, they do not exhibit similar correlation in M17-IRDC. The Spearman correlation coefficients of HCN and HCO$^{+}$ versus gas column density are 0.72 and 0.68 in  M17-H{\scriptsize II} cloud, and they are both 0.54 in M17-IRDC. 
Similar trend is also found between the integrated intensities of HCN and HCO$^{+}$  and excitation temperature except that the linear dependence of the integrated intensities of HCN and HCO$^{+}$ only appear at temperature higher than 50\,K. The Spearman correlation coefficients of HCN and HCO$^{+}$ versus excitation temperature are lower: 0.65 and 0.59 in  M17-H{\scriptsize II} cloud, and they are 0.49 and 0.39 in M17-IRDC, respectively. 
The mean ratios of HCO$^{+}$/HCN are $\sim 0.73$ in M17-H{\scriptsize II} and $\sim0.97$ in M17-IRDC. While the mean ratio in  M17-H{\scriptsize II}  is similar to values obtained for resolved nearby galaxies in a recent survey \citep{donaire19}, that of M17-IRDC is higher. Compare to Galactic Cloud, the ratio of of HCO$^{+}$/HCN in M17-H{\scriptsize II} is comparable to Ophiuchus (0.77) and higher than Aquila (0.63) but lower than Orion B (from 0.9 to 1.4) while the value of M17-IRDC is more comparable to those of Orion B \citep{shimajiri17}.

\begin{figure*}[!htbp]
\vskip -0.1cm
\hspace{-0.3cm}
 $\begin{array}{cc}
\includegraphics[width=8.5cm]{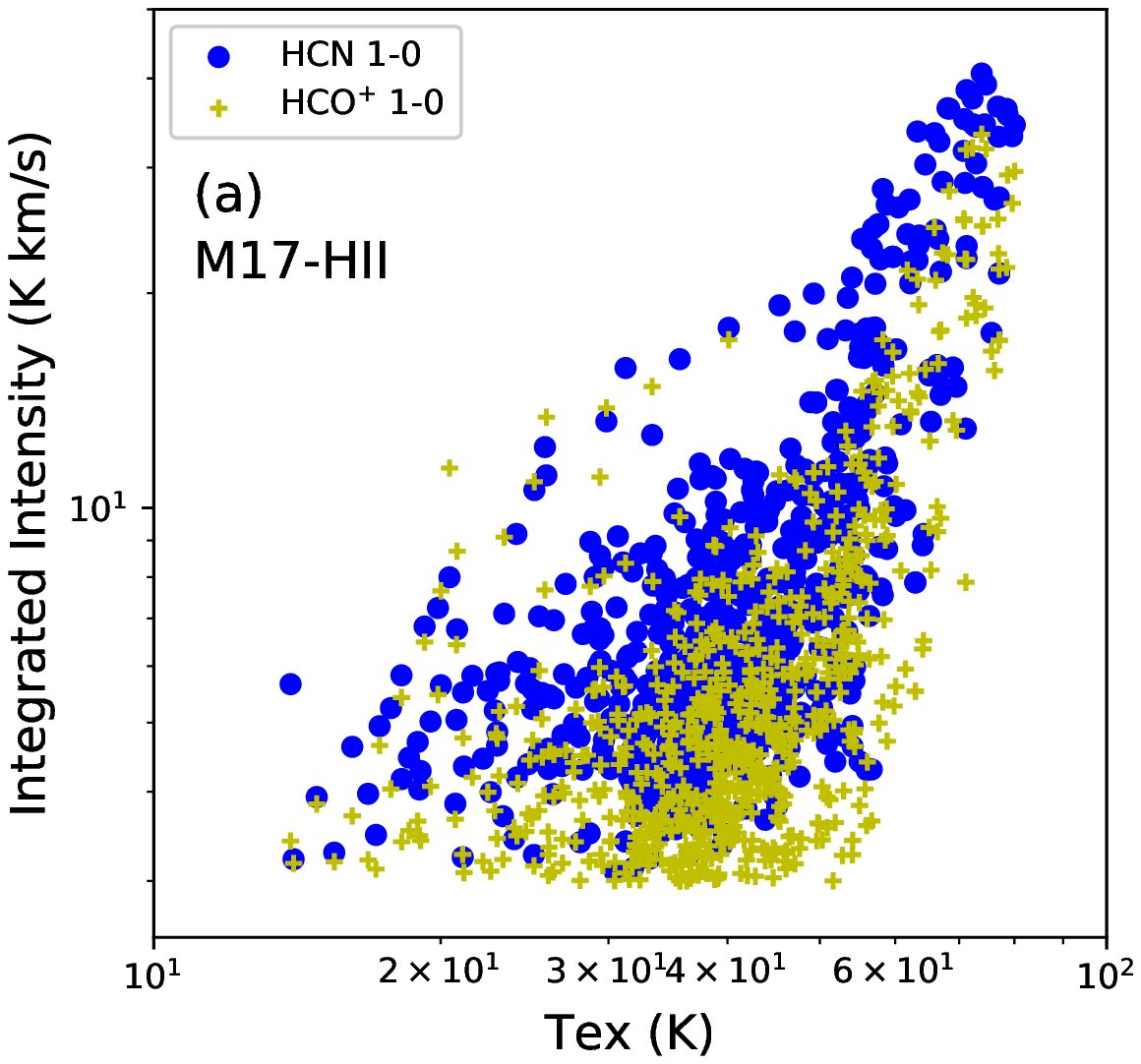} &
\includegraphics[width=9cm]{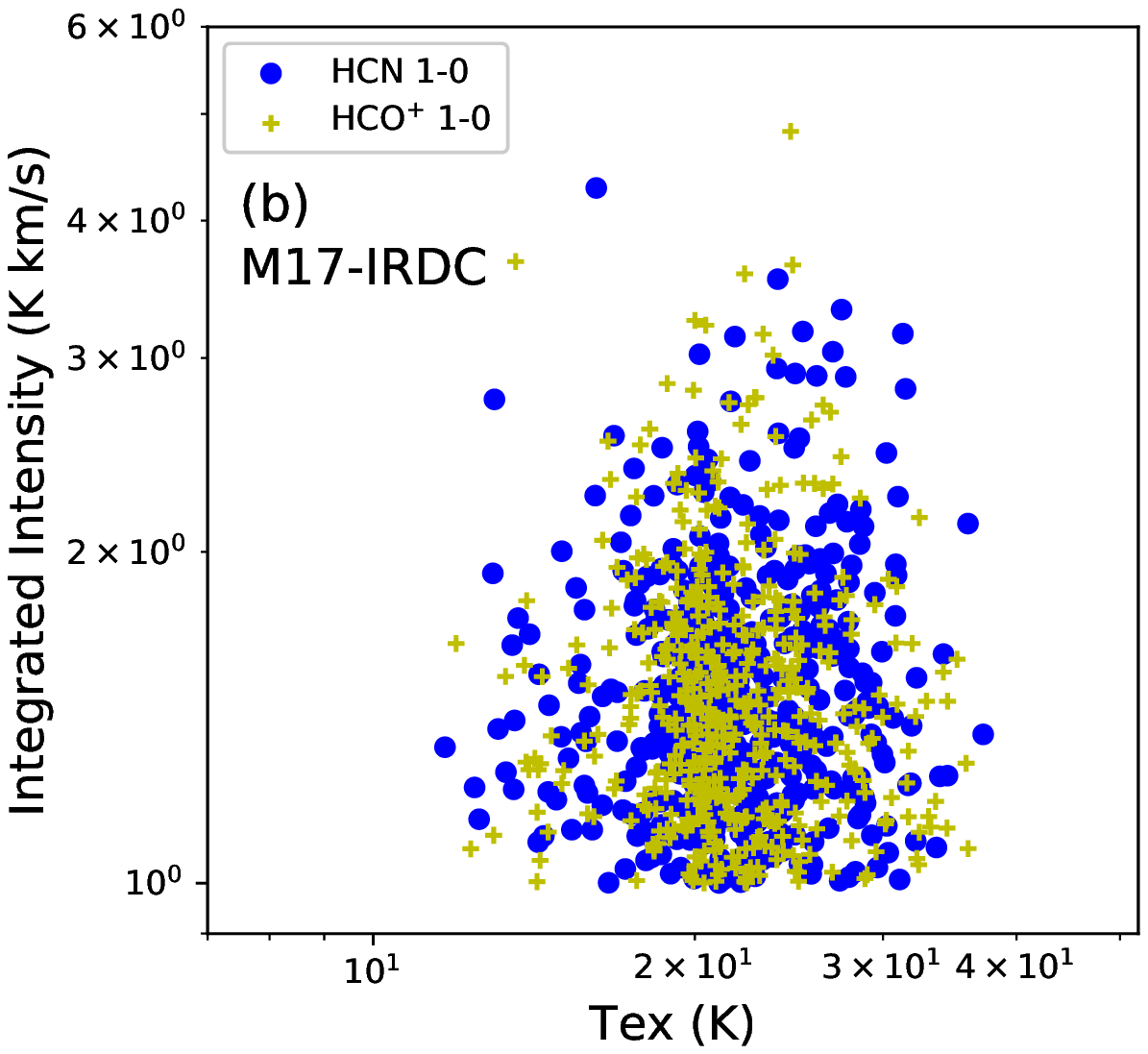} 
 \\
\hspace{-0.3cm}
\\
\end{array}$
\caption{The scatter plots of pixel-by-pixel HCN and HCO$^+$ ($J= 1-0$) integrated intensity versus gas temperature for {\bf (a)} M17-H{\scriptsize II} and {\bf (b)} M17-IRDC regions. We note that the sacle are different in these plots.}
\label{fig:m17denseT}
\end{figure*}

\section{Discussion}
\label{sect:discussion}

In the previous Section, we have quantified the difference between M17-H{\scriptsize II} and M17-IRDC in terms of the cloud mass distribution and the distribution of dense gas tracers. In this Section, we will discuss the difference between two sub-regions and how their distributions impact star formation in M17. 

\subsection{Two distinct sub-regions of in the M17 cloud complex}

\begin{figure*}[!htbp]
\vskip -0.1cm
\hspace{-0.3cm}
 $\begin{array}{c}
\includegraphics[width=18cm]{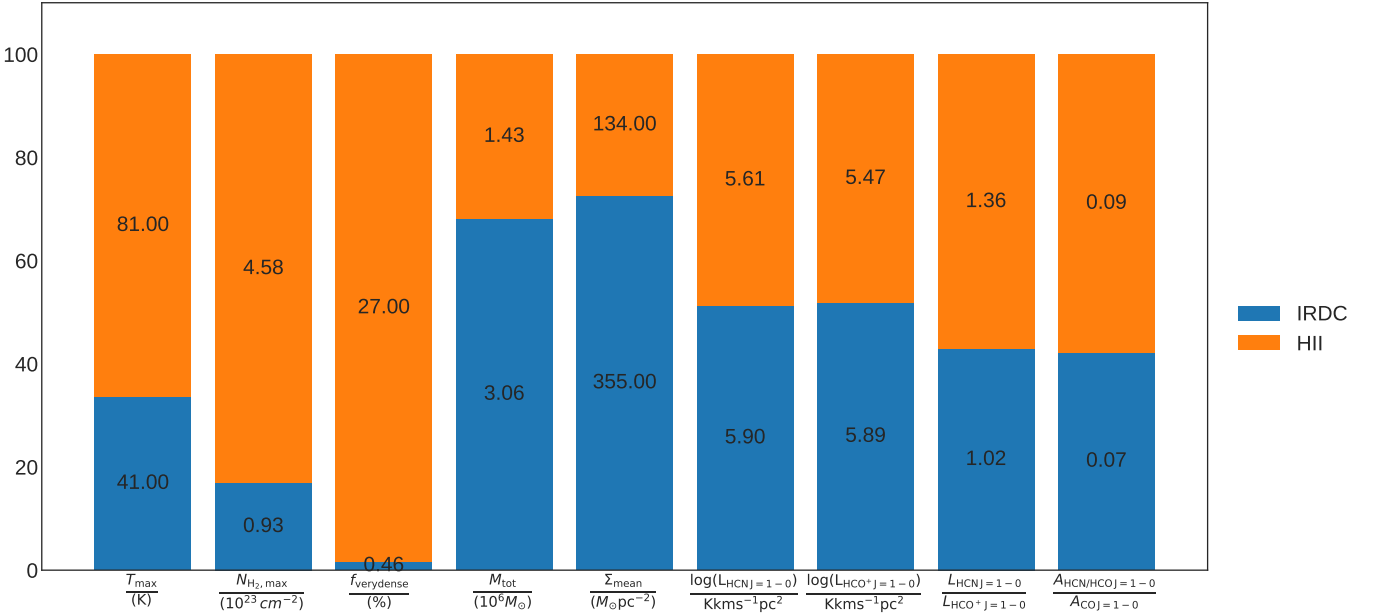} 
\hspace{-0.3cm}
\\
\end{array}$
\caption{{\nlqb A comparision of different physical properties of the M17-H{\scriptsize II} and M17-IRDC clouds: maximum excitation temperature $T_{\rm max}$, maximum column density $N_{\rm H_2, max}$, very dense gas ($N_{\rm H_2} >1\times 10^{23}$\, cm$^{-2}$) ratio $f_{\rm very dense}$,  total mass $M_{\rm tot}$, mean gas surface density $\Sigma_{\rm mean}$, HCN\,$J=1-0$ line luminosity, HCO$^+$ \,$J=1-0$ line luminosity, HCN\,$J=1-0$/HCO$^+$ \,$J=1-0$ line luminosity ratio, ratio of dense gas area over total gas area $A_{HCN or HCO^+}/A_{CO}$.}}
\label{fig:m17comparision}
\end{figure*}

M17-H{\scriptsize II} and M17-IRDC clouds have different physical properties as shown in Section\,\ref{sect:results} and are also known for having different star formation activity \citep{povich10,povich16}. We summarize the properties of the two sub-regions in Figure\,\ref{fig:m17comparision}. M17-H{\scriptsize II} is habouring a massive protostellar cluster  \citep{kuhn15} while M17-IRDC is quiescient and currently forming low-mass stars \citep{ohashi16}. M17-H{\scriptsize II} is warmer and has higher peak column density and much higher very dense gas fraction. M17-IRDC is more massive and has higher mean surface density. The total fraction of $\frac{L_{HCN\,J=1-0}}{L_{HCO^+ \,J=1-0}}$ and fraction of dense gas area over CO gas area is also higher in M17-H{\scriptsize II} (see Section\,\ref{sect:densegasproxy}).

\subsection{HCO$^{+}$ and HCN (J = 1--0) as proxies to trace dense gas mass}
\label{sect:densegasproxy}
\begin{figure*}[!htbp]
\vskip -0.1cm
 $\begin{array}{cc}
\includegraphics[width=8.5cm]{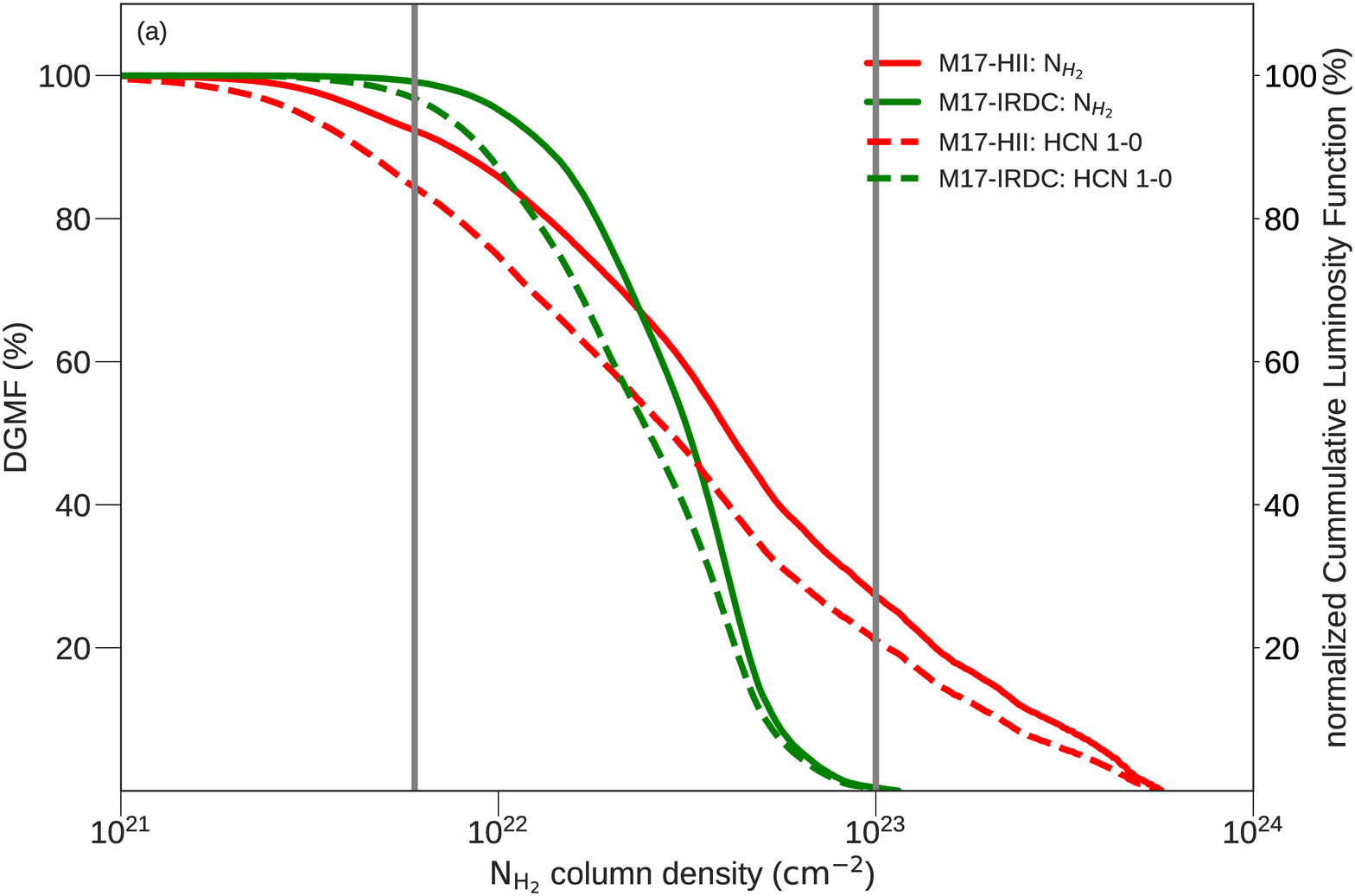}&
\includegraphics[width=8.5cm]{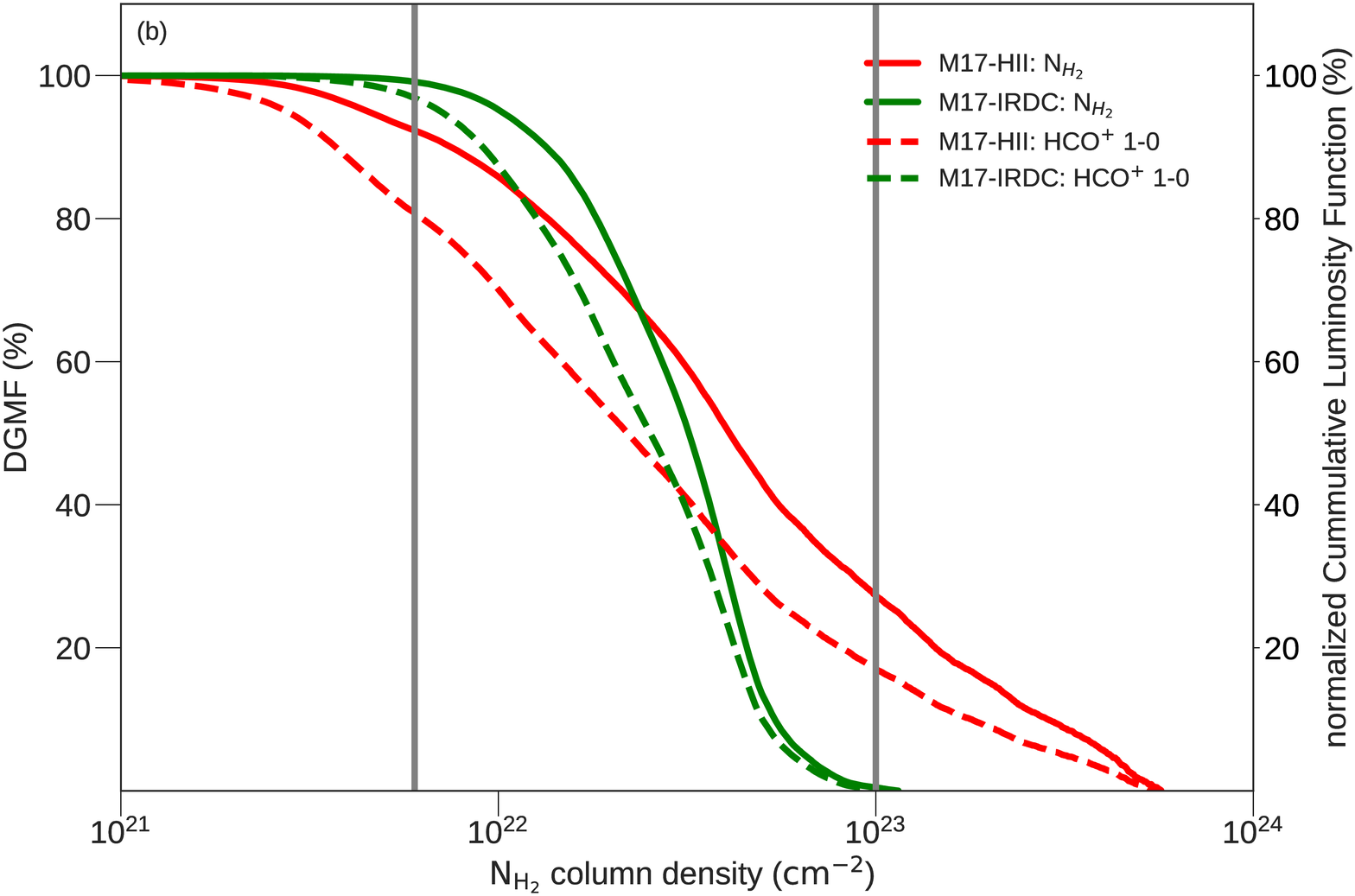}\\
  
\hspace{-0.3cm}
\\
\end{array}$
\caption{The HCN {\bf (a)} and HCO$^+$ ($J=1-0$) {\bf (b)} normalized cumulative luminosity as compared to the gas column density CMD.
}
\label{fig:m17L}
\end{figure*}

The distributions of HCO$^{+}$ and HCN ($J=1-0$) coincide with the high density part of the gas column density map, suggesting that HCO$^{+}$ and HCN ($J=1-0$) emission lines trace the similar density region.  
As in Section\,\ref{sect:cummilative}, we divide the integrated intensity maps of HCN and HCO$^{+}$ ($J=1-0$) into 120 bins based on the gas column density bins ranging from $10^{21}$\,\cmd\, to $10^{24}$\,\cmd . We then calculate the line luminosity function (LLF) in the formed of the normalized cumulative line luminosity function (nCLLF) as:
\begin{equation}
{\rm LLF(>N_{\rm H_2})} =\frac{\int_{N_{\rm H_2}}^{N_{\rm H_2,max}} A({N_{\rm H_2}})   dL_{\rm line}(N_{\rm H_2})}{\int_{0}^{N_{\rm H_2, max}} A({N_{\rm H_2}})   dL_{\rm line}(N_{\rm H_2})} \ .
\label{eq:CMD}
\end{equation}
{\nlq We use the same approach detailed in \cite{solomon97} or \cite{wu10} to calculate the line luminosity as
\begin{equation}
L_{\rm line} = 23.5\Omega_{s*b}D_{\rm L}^{2}I_{\rm line} \left(1+z\right)^{-3}
\end{equation} 
where $\Omega_{s*b}$ is the source solid angle convolved with the telescope beam and has a unit of arcsec$^{2}$. $D_{\rm L}$ is the luminosity distance and has a unit of Mpc, $I_{\rm line}$ is the integrated line intensity and has a unit of K $\kms$. z is the redshift of the object. The luminosity is expressed in the unit of K\kms pc$^{2}$. If the source is much smaller than the beam, then  $\Omega_{s*b}$ is equal to the beam size  $\Omega_{b}$. In the case of Galactic clumps and clouds, the sources are often more extended than the beam, the source solid angle convolved with a Gaussian beam is expressed as 
\begin{equation}
\Omega_{s*b} = \left( \frac{\pi\times\theta_{s}^{2}}{4ln2} \right) 
\left( \frac{\theta_{s}^{2}+\theta_{\rm beam}^{2}}{\theta_{s}^{2}}\right).
\end{equation}
Therefore, the line luminosity in unit of K\kms pc$^{2}$ of an object in the Milky Way ($z=0$) can be derived as 
\begin{equation}
L_{\rm line} = 23.5\times 10^{-6} \Omega_{s*b}d^{2}I_{\rm line}
\end{equation}
where $d$ is distance and has a unit of kpc and $T_{\rm MB}$ is the main-beam brightness temperature of the line. The line luminosity can be converted from  K\kms pc$^{2}$ to solar bolometric luminosity unit as in \cite{nguyenluong13}.
The line luminosity function-$N_{\rm H_2}$ profiles are shown in Figure \ref{fig:m17L} and are also compared with DGMFs.}

The luminosity functions of both lines span the entire column density range traced by the gas column density and are shallower than the DGMF profiles. Therefore, both HCO$^{+}$ and HCN ($J=1-0$) integrated intensities can serve as  proxies for high column density dense gas tracers, at least at the scale smaller than cloud scale. {\nlq In contrast, other works have shown that HCN can trace more diffuse gas than HCO$^{+}$ (i.e. \citealt{kauffmann17,pety17,shimajiri17}). The difference comes from the fact that M17 is a massive star-forming regions while the molecular clouds in other studies are low-mass forming regions. }

While the luminosity functions of HCO$^{+}$ and HCN are similar in M17-IRDC, the HCN luminosity function is higher than that of HCO$^{+}$ in M17-H{\scriptsize II}. Therefore, 
the ratio of HCN/HCO$^{+}$ in M17-IRDC is lower than the ratio in M17-H{\scriptsize II}. Lower HCN/HCO$^{+}$ might be related to the quiescent state of on-going state of the IRDC while the higher ratio of HCN/HCO$^{+}$ might reflect the cloud at a more advanced state as free electrons in the turbulent and Far-UV irradiated environment near the massive cluster can easily recombined with HCO$^{+}$ \citep{papadopoulos07}.  It agrees with the fact that HCN is enhanced in X-Ray dominated environment that is produced by young massive stars in M17-H{\scriptsize II}. Actually, \citet{meijerink05} and \citet{meijerink07} show that a lower ratio is observed on the surface of the X-ray dominated region or photon-dominated region and a higher ratio is often seen in high density and cold regions. {\nlq Therefore, a high HCN/HCO$^{+}$ ratio might be a good indicator of massive star formation or more evolved evolutionary stages of star formation (see also \citealt{wu10,sanhueza12}). }
\subsection{Stability of clumps in M17-IRDC and M17-H{\scriptsize II}}
\label{sect:stability}

\begin{figure*}[!htbp]
\vskip -0.1cm
\hspace{-0.3cm}
 $\begin{array}{c}
\includegraphics[width=17cm]{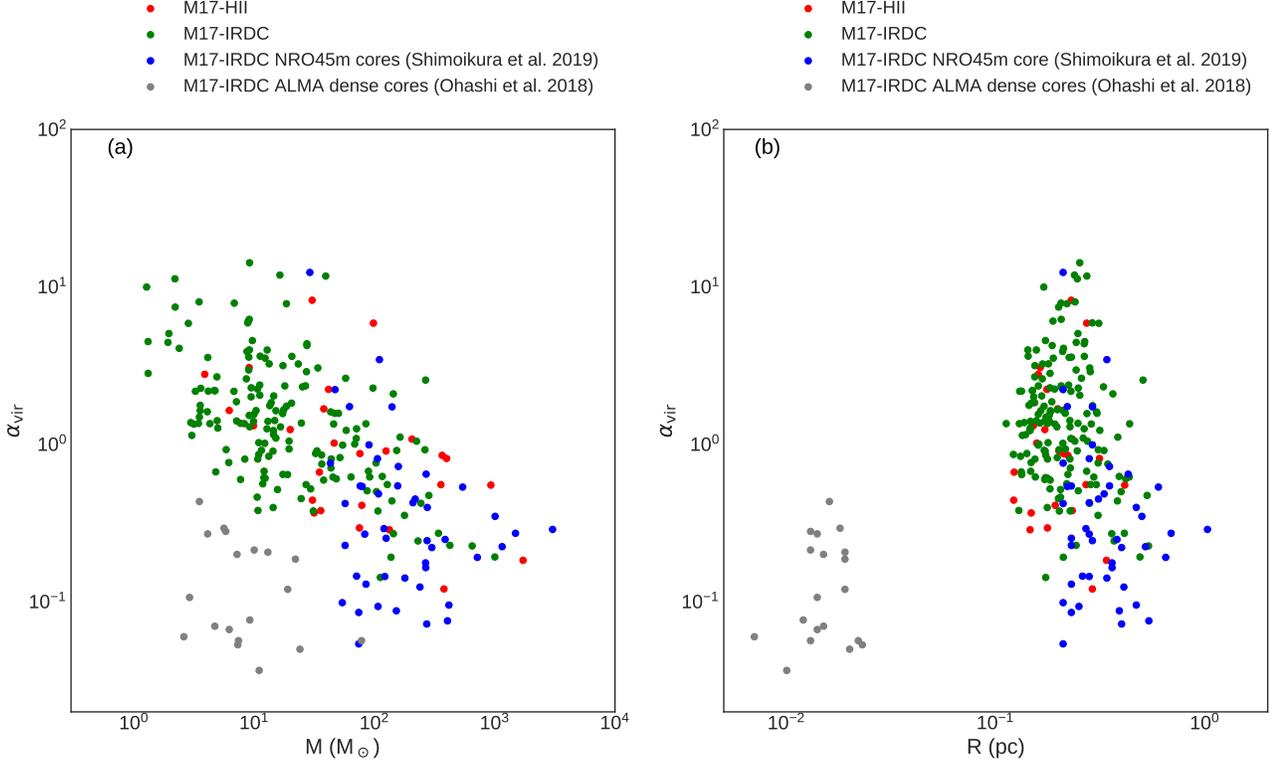}
\hspace{-0.3cm}
\\
\end{array}$
\caption{Virial parameters as functions of (a) mass and (b) 
radius of individual clumpss in M17-H{\scriptsize II} (red) and M17-IRDC regions (green).  Virial parameters of N$_{2}$H$^{+}$  (J = 1--0) cores (blue) derived from NRO 45m observations \citep{shimoikura19a} and N$_{2}$H$^{+}$  (J = 1--0) dense cores (grey) derived from ALMA observations \citep{ohashi16} are also plotted, respectively. }
\label{fig:viriallog}
\end{figure*}

There are more unbound clumps in M17-IRDC
region than in M17-H{\scriptsize II} region, as seen in the $\alpha_{\rm vir}-M$ and $\alpha_{\rm vir}-R$ relations
in a log scale (Figure\,\ref{fig:viriallog}). 64\% of the clumps in M17-IRDC have $\alpha_{vir} >1$.
In addition, the clumps in M17-IRDC have systematically higher virial
parameters than those in M17-H{\scriptsize II}. These facts support the idea that the clumps in
M17-H{\scriptsize II} are more prone to gravitational contraction while those in M17-IRDC that have  $\alpha_{vir} >1$ are gravitationally unbound and dispersing. 
However, there is a chance that they can be in the gravitational equilibrium if surrounded by the high external pressure (see \citealt{shimoikura19a}).

\begin{figure}[!htbp]
\vskip -0.1cm
\hspace{-0.3cm}
 $\begin{array}{c}
\includegraphics[width=8cm]{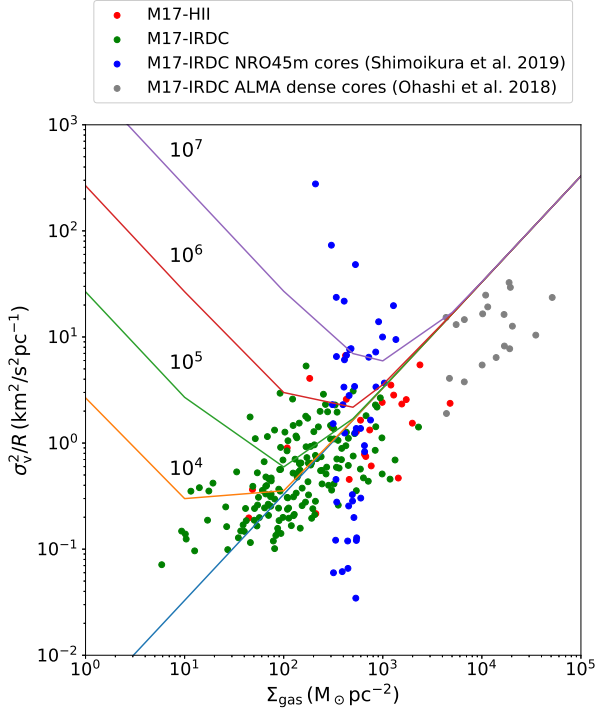} 
\hspace{-0.3cm}
\\
\end{array}$
\caption{The linewidth-size scaling coeffient and mass surface density relation. The blue straight line show the case where there is no external pressure. The orange, green, red, pink curves show the cases where extermal pressure $P_{\rm e}/k$ are 10$^{4}$, 10$^{5}$, 10$^{6}$, and 10$^{7}$ K cm$^{-3}$, respectively. 
}
\label{fig:stability}
\end{figure}

We compare the virial parameters of clumps in our studies with those of  N$_{2}$H$^{+}$  (J = 1--0) cores (radius $\sim 0.1 - 1$ pc ) obtained from observation also with NRO 45m \citep{shimoikura19a} and N$_{2}$H$^{+}$  (J = 1--0) dense cores (radius $\sim 0.007 - 0.03$ pc ) obtained from observation with the ALMA interferometer \citep{ohashi16}.  For the first dataset, we derive the 3D virial parameter using their virial masses and core masses by applying Equation~\ref{eq:virial}. For the second one, {\nlq because their virial parameters is calculated as $\frac{5\sigma^{2}R}{GM}$,} we divide it by 3 to obtain the 3D virial parameters.  While all  ALMA dense cores have $\alpha_{\rm vir} < 1$, most of NRO 45m cores have $\alpha_{\rm vir} < 1$ and some cores have $\alpha_{\rm vir} >1$. The trend is consistent with the general suggestion that virial parameter increase with size \citep{ohashi16,chen19}. We suggest that virial parameter increase with size until the core size reach $\sim0.5-1$ pc and decrease with size starting at the clump scales. We note that the 3D virial parameters of molecular cloud complex (radius $\sim 50-100$~pc) are in the range of 0.5--3 \citep{nguyenluong16}.
 
{\nlq We also examine the significance of external pressures by plotting the relationship between $\sigma_{\rm 3D}^{2}/R$ and the mass surface density $\Sigma_{gas}$ in Figure\,\ref{fig:stability} as theoretically suggested by \cite{field11} as:
\begin{equation}
\frac{\sigma_{\rm 3D}^{2}}{R} =\frac{1}{3}\left(\pi\Gamma G\Sigma + \frac{4P_{\rm e}}{\Sigma}  \right)
\label{eq:pe1}
\end{equation}
where $\Gamma$ is a form factor which equals to 0.73 for an isothermal sphere of critical mass with a centrally concentration internal density structure \citep{elmegreen89}. $P_{\rm e}$ is the external pressure and is often expressed as $\frac{P_{\rm e}}{k}$ in unit of K $\rm cm^{-3}$. 


When $P_{\rm e}$ equals 0, the cloud is in simple virial equilibrium (SVE) with the internal kinetic energy being equal to half the gravitational energy \citep{field11}. In Figure\,\ref{fig:stability}, we overlaid the simple virial equilibrium line in addition to some curves with external pressure ranging from $10^{3}-10^{7}\,\rm K\, cm^{-3}$. Contrary to the clumps in \cite{heyer09} that mostly lie above the SVE line, our data show that half of the clouds are above and half are below the SVE line. The first half is dynamically unstable due to external pressure while the second half is gravitationally bound and collapses on its own gravity. This is an interesting fact since most of the high resolution and high density cores \citep{ohashi16,shimoikura19a} can collapse on its own while our CO clumps need external pressure to collapse. Another interesting fact is that most clumps in M17-IRDC live above the SVE line which mean they need additional external pressure to collapse while those in M17-H{\scriptsize II} do not. This fact is consistent with the higher virial parameters in M17-IRDC clumps and also consistent with what found in molecular clouds in the Galactic Center \citep{miura18}.}
\subsection{Conditions for massive star formation in M17 H{\scriptsize II} and M17-IRDC}
\label{sect:conditions}


The mass functions of YSOs in the M17-H{\scriptsize II} and M17-IRDC regions were derived by \citet{povich10} and \citet{povich16} using Spitzer data. They found that the high-mass stellar population is deficient in M17-IRDC, 
claiming a possibility that the massive star formation is delayed in the M17-IRDC region assuming that the final stellar IMF approaches to the Salpeter IMF. Here, we attempt to elucidate whether the present physical conditions 
of M17 satisfy the criteria of massive star formation,
based on our observational results. 
\begin{figure*}[!htbp]
\vskip -0.1cm
\hspace{-0.3cm}
 $\begin{array}{c}
\includegraphics[width=18.5cm]{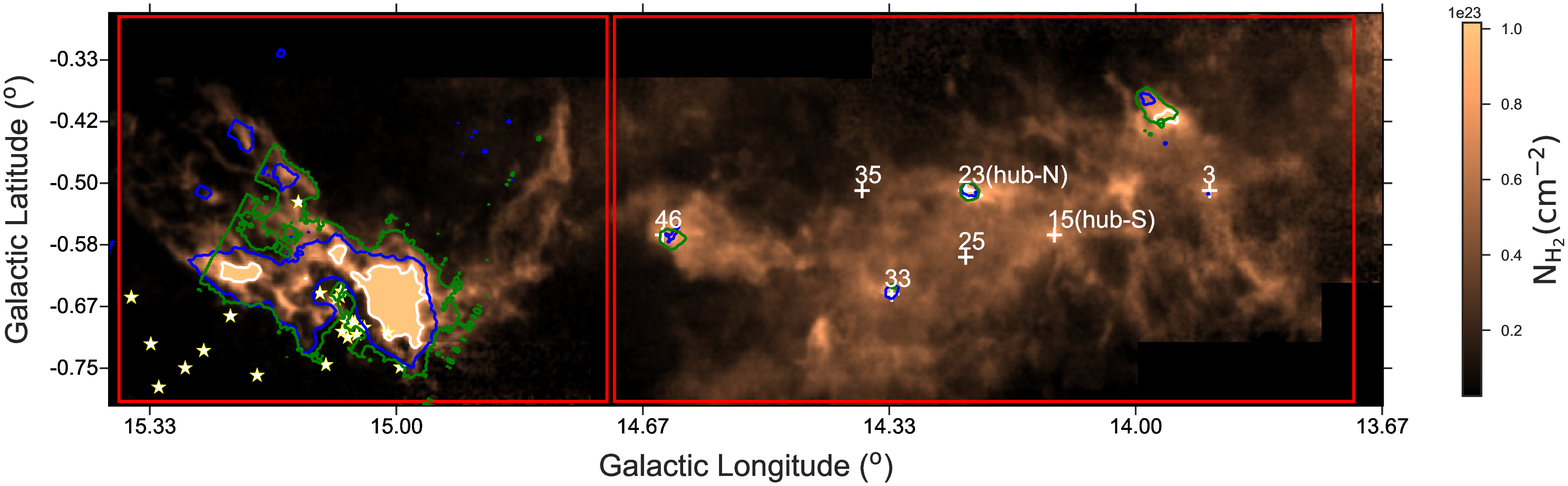} \\
\hspace{-0.3cm}
\\
\end{array}$
\caption{Gas column density maps and dense gas distribution. The coloured scale image indicates the column density distribution derived from $^{13}$CO. The white contour is at the column density of $\sim$ 1 g cm$^{-2}$. Blue contours represent the HCO$^{+}$ integrated intensity and the green contours represent the HCN integrated intensity, both at 3.5 K km/s level.  {The star symbols pinpoint the OB star clusters responsible for the giant H{\scriptsize II} region.} The cross symbols pinpoint the locations of the massive cores (M $>$ 500\,\msun) reported by  \cite{shimoikura19a}.}
\label{fig:overdense}
\end{figure*}
\citet{krumholz08} proposed the threshold column density as one condition for massive star formation of
$\Sigma > 1$ g cm$^{-2}$ or $\rm N_{\rm H_2}  > 10^{23}\,{\rm cm}^{-2}$, beyond which molecular clouds could form high-mass stars, avoiding 
further fragmentation to form lower-mass cores. 
As shown in Section \ref{sect:results}, the cumulative column density distributions derived from our CO data
(see Figure \ref{fig:DGMF}) clearly indicate that M17-H{\scriptsize II} region is denser 
than the M17-IRDC. In addition, M17-H{\scriptsize II} region is more evolved in term of forming stars and M17-IRDC region is more quiescent. Although  
the median column density and total mass of M17-IRDC are larger than those of the M17-H{\scriptsize II} region, 
most of the clumps identified in M17-IRDC have column densities lower than the threshold of massive star formation. 
In contrast, M17-H{\scriptsize II} contains several clumps having column densities comparable to or larger than the threshold.
For comparison, we show the dense region in M17 traced by HCO$^{+}$ and HCN  $J=(1-0)$ emission in Figure \ref{fig:overdense}.
Therefore, we conclude that the current mass concentration in M17-IRDC is not enough to efficiently create high-mass stars. However, there is a large mass reservoir and a filamentary network that can feed the clumps and cores to grow up and in the future to form high-mass stars \citep{ohashi16}. {\nlq External pressure might be needed for the clumps in M17-IRDC to collapse and form stars as shown in Section\,\ref{sect:stability}. }

\subsection{What can be the star formation scenario for the entire M17 cloud complex?}
\label{sect:SF}
In Section \ref{sect:results}, we show that the dense gas fraction and the dynamical states of the M17-H{\scriptsize II} cloud and clumps within it are different from those of M17-IRDC. And yet, both regions were suggested to be connected \citep{shimoikura19a} and are formed by colliding clouds (M17-H{\scriptsize II}: \citealt{nishimura18} and M17-IRDC: \citealt{sugitani19}).
In addition, \cite{sugitani19} used the the near-infrared polarization and CO ($J=3-2$) data to discover that the main elongation axis of M17-IRDC is perpendicular to the global magnetic fields, which are roughly perpendicular to the Galactic plane.  
Such a structural and magnetic configuration might be consistent with the molecular cloud formation by the Parker instability \citep{parker66,shibata92}.
Since the Parker instability is more unstable to the anti-symmetric mode, the field lines tend to cross the Galactic plane. According to the linear stability analysis of the gravitational instability of magnetized Galactic disks \citep{hanawa92}, the gravitational instability is unstable only for the symmetric mode (see also \citealt{nakamura91}).
When the GMCs formed by the Parker instability, large-scale colliding clouds are a natural outcome because the clouds slide down along the field lines and accumulated in the valleys of magnetic fields.

This mechanism is also consistent with the suggestion of \cite{elmegreen79} who suggested that the Galactic spiral density wave
passed through the whole M17 region from north-east to south-west, triggering the star formation 
in M17-H{\scriptsize II}. In observations, large-scale colliding flows on the large scale is proven to be an effective way to convert atomic gas to molecular gas and from diffuse gas to dense gas \citep{nguyenluong11b,nguyenluong11,nguyenluong13,motte14}. See also the review on massive star formation \citep{motte18}. \cite{sugitani19} also pointed out that two gas components with velocities of $\sim$20 \kms (main)  and $\sim$35 \kms (secondary) in M17-IRDC could collide. These components can form a broad feature bridge, which is a collection of diffuse gas in that has system velocity in between $\sim$20 \kms (main)  and $\sim$35 \kms (\citealt{sugitani19}, Kinoshita et al. in prep.) that is a signature of cloud-cloud-collision as suggested by numerical simulation of gas kinematics \citep{haworth15,haworth15b,inoue13} and by gas kinematic observations \citep{torii11,torii15,torii17,torii18,torii18b,fukui14,fukui16,fukui18,fukui18b,fujita19,tsuboi15}.


%
%

%
\section{Conclusion}
\label{sect:conclusion}
We examined the molecular gas structure of the M17 complex using $^{12}$CO~($J=1-0$), $^{13}$CO~($J=1-0$), HCO$^+$ ($J=1-0$), HCN ($J=1-0$) and other supplement data. Our main findings are summarized as follows:

\begin{enumerate}
\item  There are three main molecular clouds with systemic velocities $\sim$ 20 \kms, 38 \kms, and 57 \kms\, along the line-of-sight of M17 complex.
The main component of 20 \kms\, sits in the Sagittarius arm at a distance of $\sim$ 2 kpc. The 38 \kms\, and 57 \kms\, clouds might be associated with Scutum and Norma arms, respectively.
\item  The cloud complex can be divided into two clouds: M17-H{\scriptsize II} region containing the NGC~6618 star cluster and M17-IRDC region containing the prominent IRDC network.  
We confirmed from the molecular observations that M17-H{\scriptsize II} has a significant fraction of molecular gas (27\%) with high column densities and high volume densities that surpasses massive star formation threshold of 1 g cm$^{-2}$, whereas such high column densities and high volume densities gas is deficient in M17-IRDC (only 0.46\%).
\item M17-H{\scriptsize II} has a higher dense gas fraction than M17-IRDC as seen in the Cumulative Mass Distributions of gas column density and the normalized cumulative line luminosity of dense gas tracers such as HCN and HCO$^{+}$ ($J=1-0$). Observations of HCO$^+$ and HCN emission also confirmed that more denser gas presents in M17-{\scriptsize II} region than in M17-IRDC.
\item {\nlq HCO$^+$ and HCN emission trace all gas with column density higher than $3\times 10^{22}$ cm$^{-2}$} and 
higher HCN/HCO$^{+}$ ratio in M17-{\scriptsize II} region might be a good indicator of massive star formation or more evolved evolutionary stages of star formation in M17-{\scriptsize II}.

\item Applying a Dendrogram analysis to the $^{12}$CO data, we identified clumps in the two clouds.  Clumps in M17-H{\scriptsize II} are more compact than those in M17-IRDC and most clumps have virial parameters $\alpha_{\rm vir} <=1$. 
On the other hand, most clumps in M17-IRDC have virial parameters $\alpha_{\rm vir} >1$. {\nlq Clumps in M17-IRDC need external pressure while clumps in 17-H{\scriptsize II} can collapse under their own gravity.}
Such distinct dynamical states of clumps is consistent with the current activity of star formation where M17-H{\scriptsize II} is forming massive stars more efficient in the past while intense massive star formation is not happening yet in M17-IRDC.
\item {\nlq M17 complex appears to have been formed as a whole by large scale compression. This compression triggered star formation in M17-H{\scriptsize II} and would also trigger star formation in M17-IRDC in the future.}

\end{enumerate}

\acknowledgments
A part of this work was financially supported by JSPS KAKENHI Grant Numbers
JP17H02863, JP17H01118, JP26287030, and JP17K00963. 
The 45-m radio telescope is operated by NRO, a branch of NAOJ.

\bibliographystyle{aasjournal}

\appendix

\section{Dendrogram masked map and results}

\begin{figure*}[!hbp]
 \begin{center}
 \includegraphics[width=18.cm]{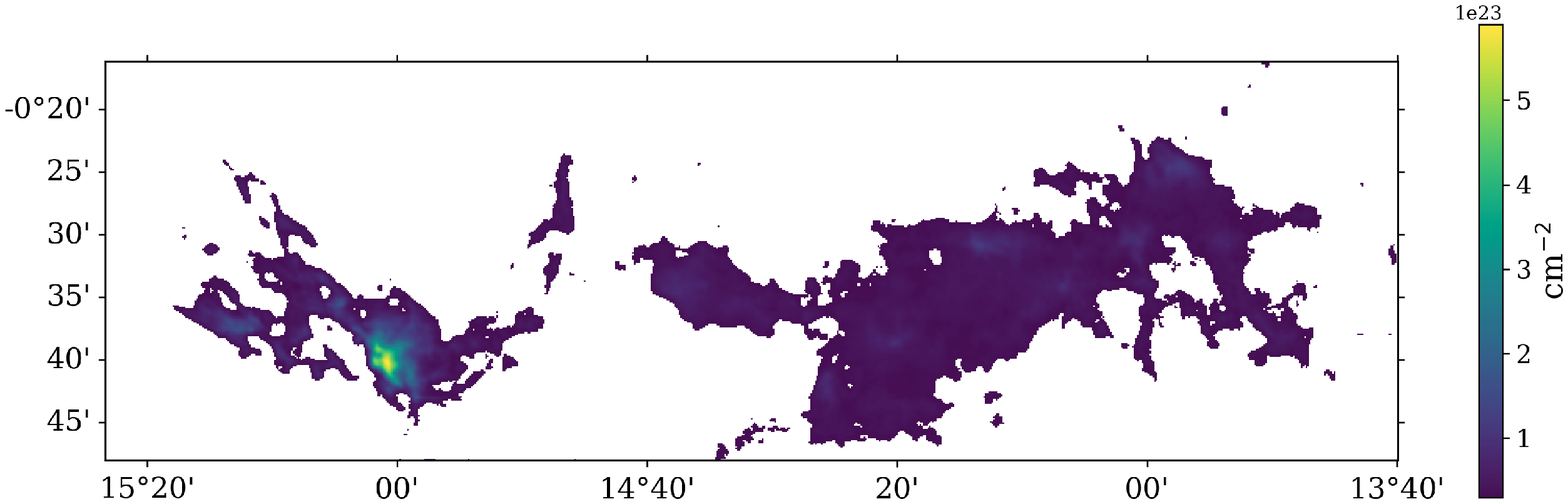}\\
 \vskip 1cm
 
 \includegraphics[width=18.cm]{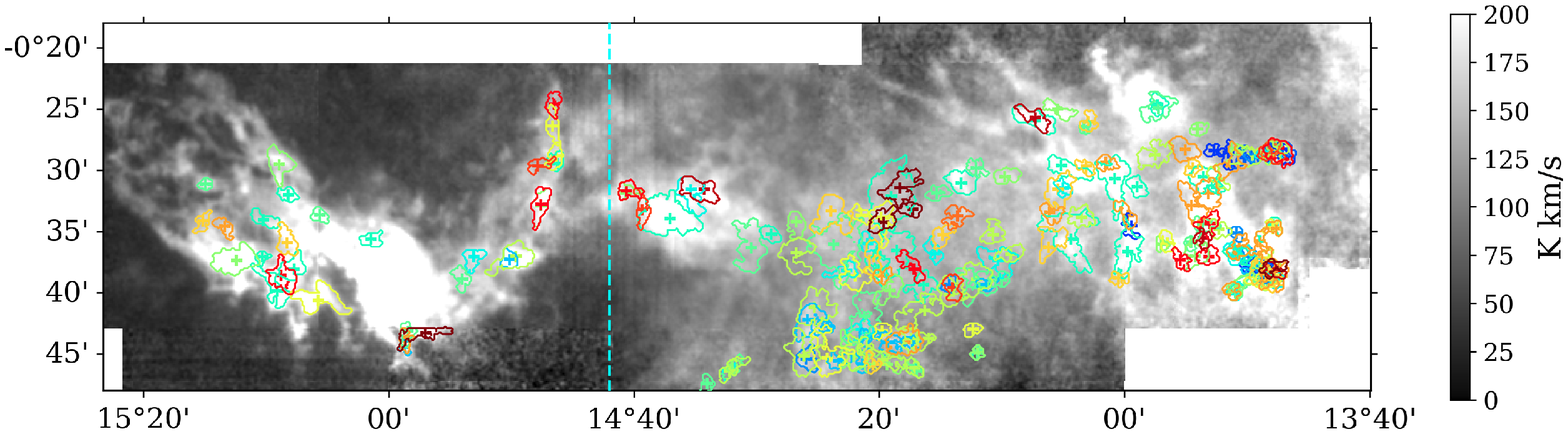}
 \end{center}
\caption{The masked map used to extract sources with Dendrogram where we masked all pixels with column density below $3\times10^{22}$\,cm$^{-2}$ {\bf (upper)} and the detected leaf structures extracted with the Dendrogram {\bf (lower)}. }
\label{fig:Dendrogram}
\end{figure*}

Figure \,\ref{fig:Dendrogram}a shows the masked map used to extract sources with Dendrogram where we masked out all pixels with column density below $3\times10^{22}$\,cm$^{-2}$ and Figure \,\ref{fig:Dendrogram}b shows the detected leaf structures extracted with Dendrogram. 

\section{$^{12}$CO\, and $^{13}$CO\, ($J=1-0$) intensity maps of the clouds along the line of sight of M17 complex integrated in different velocity ranges }
\begin{figure*}[!hbp]
 \begin{center}
 \includegraphics[width=16cm]{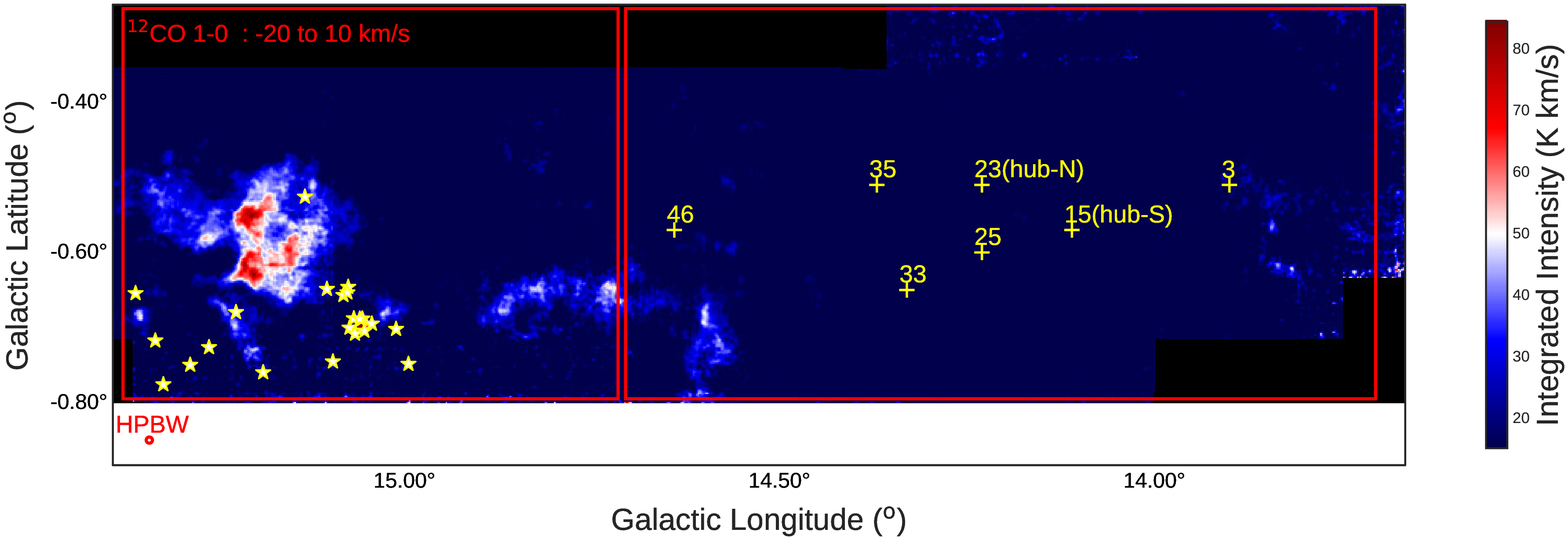}\\
  \includegraphics[width=16cm]{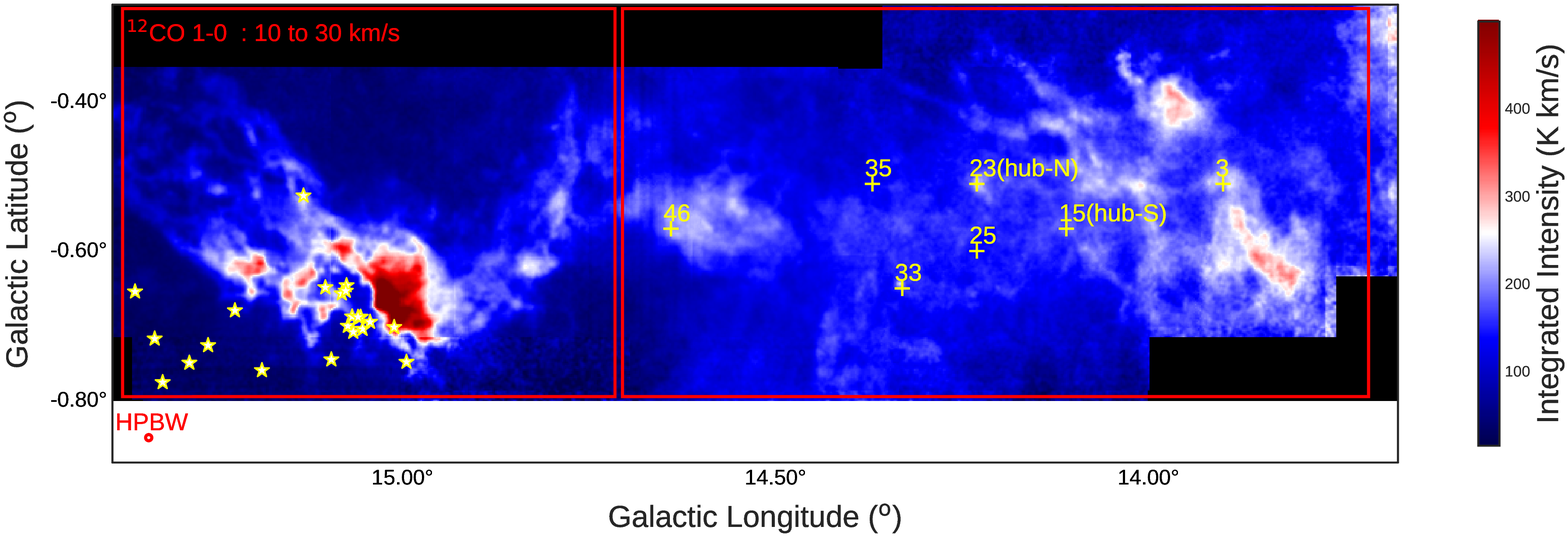}\\
    \includegraphics[width=16cm]{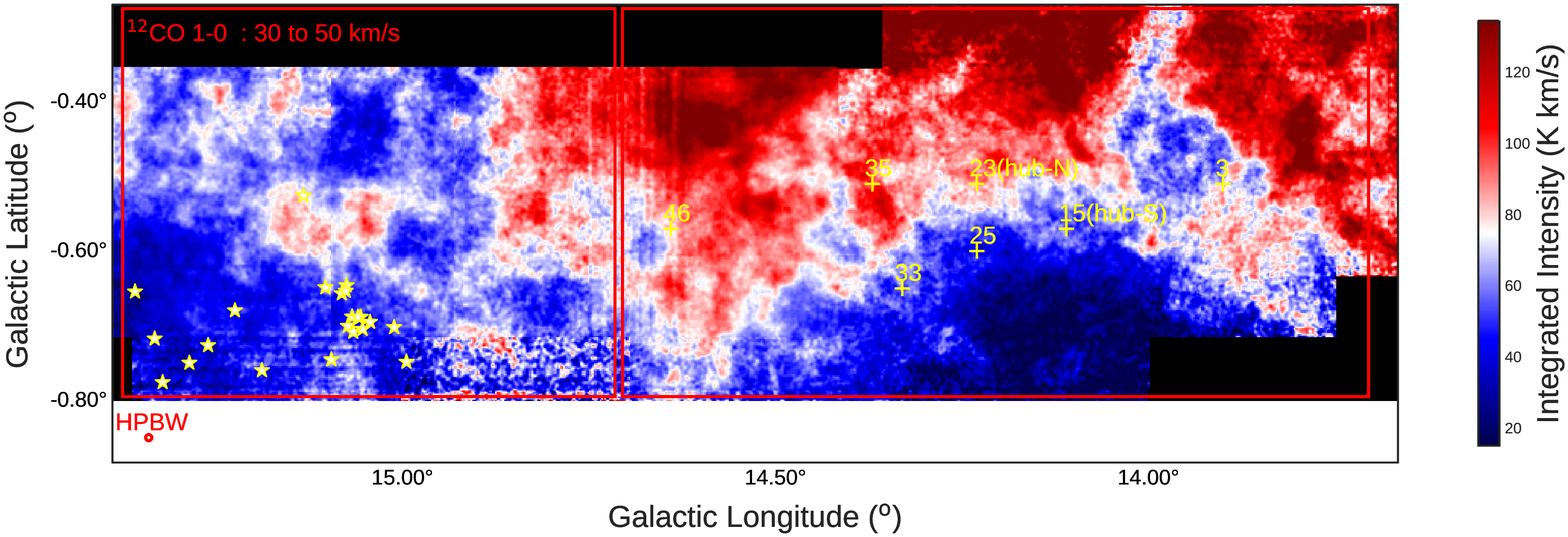}\\
    \includegraphics[width=16cm]{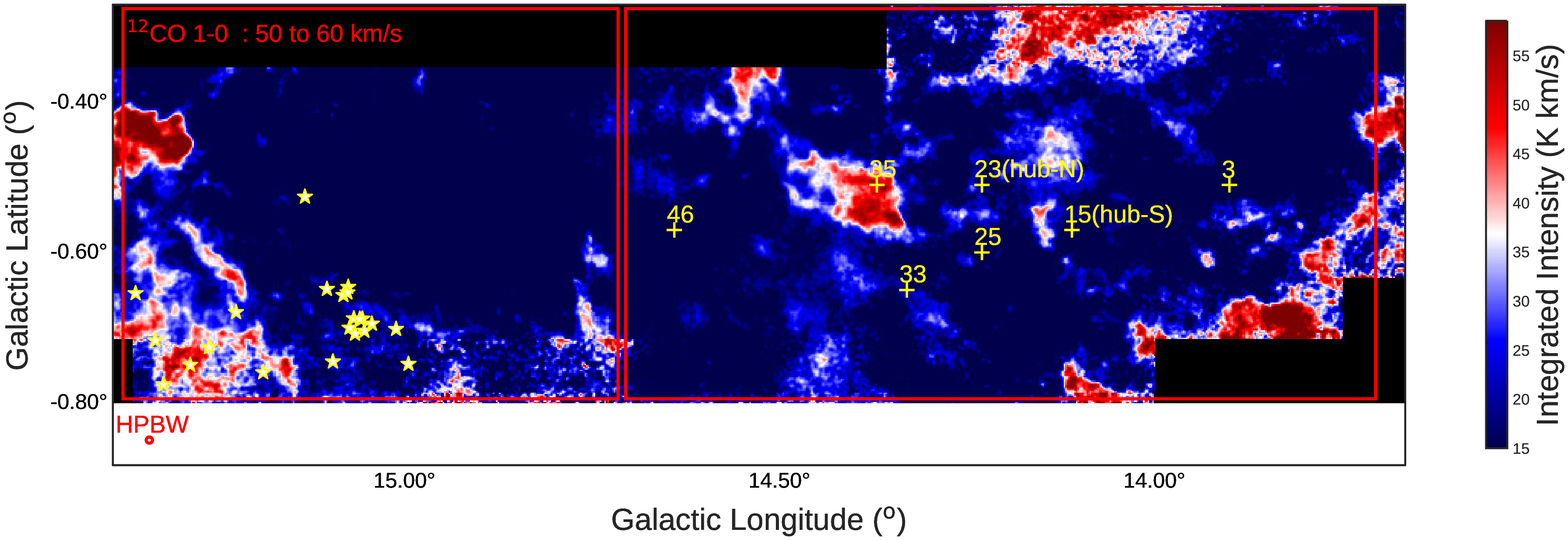}
 \end{center}
\caption{$^{12}$CO\,($J=1-0$) integrated intensity maps of the M17 complex. The velocity range for integration is indicated in the upper-left corner of each panel. 
The red rectangles outline the two prominent star-forming regions: M17-H{\scriptsize II} (left) and M17-IRDC (right).  {The star symbols pinpoint the OB star clusters responsible for the giant H{\scriptsize II} region.} The cross symbols pinpoint the locations of the massive cores (M $>$ 500\,\msun) reported by \cite{shimoikura19a}.}
\label{fig:m17_12co_intall}
\end{figure*}

\begin{figure*}
 \begin{center}
 \includegraphics[width=16.cm]{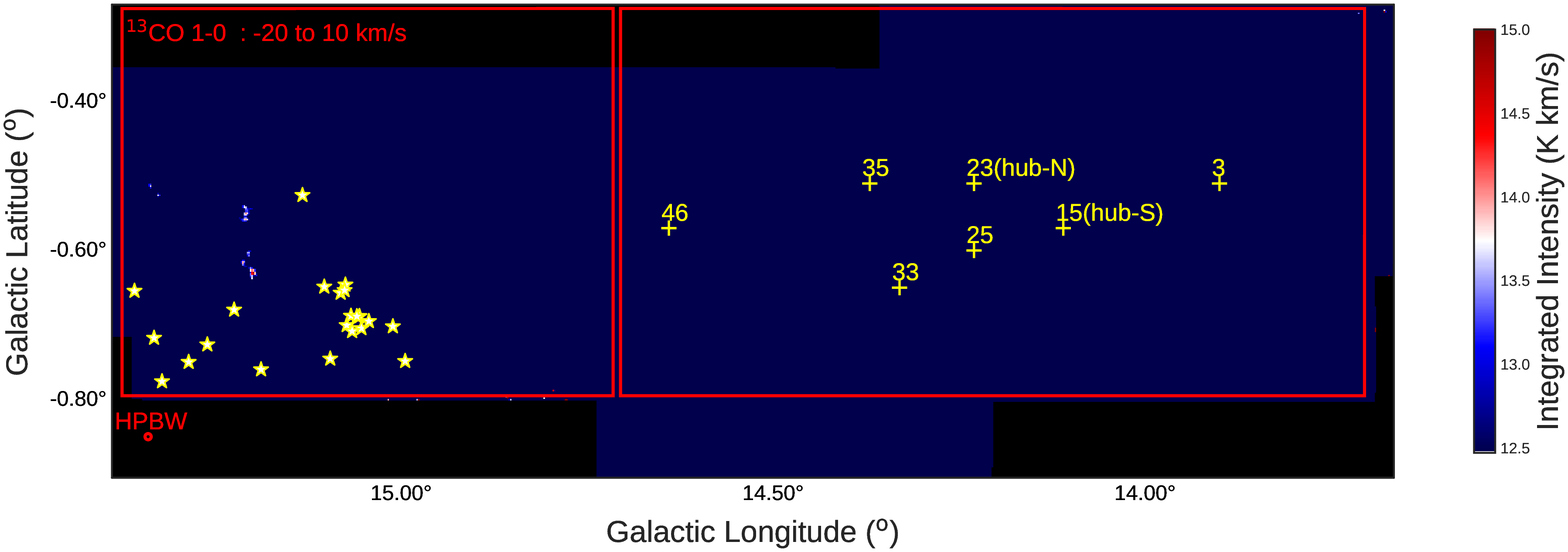}\\
 \includegraphics[width=16.cm]{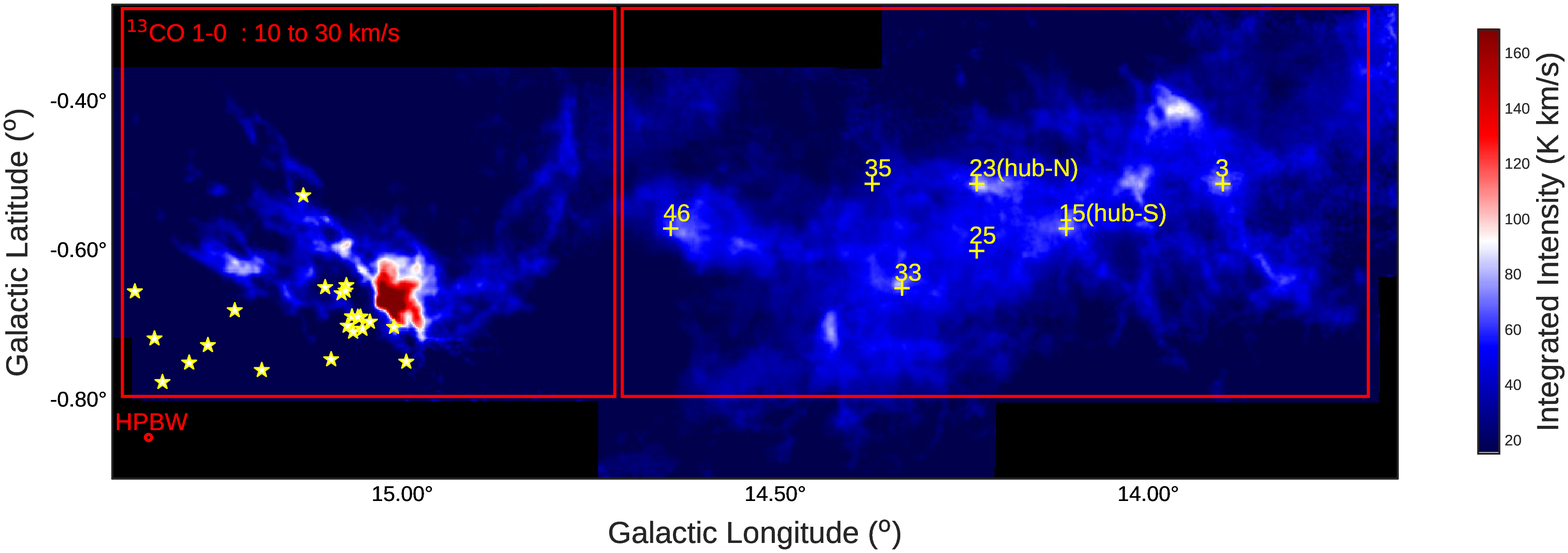}\\
    \includegraphics[width=16.cm]{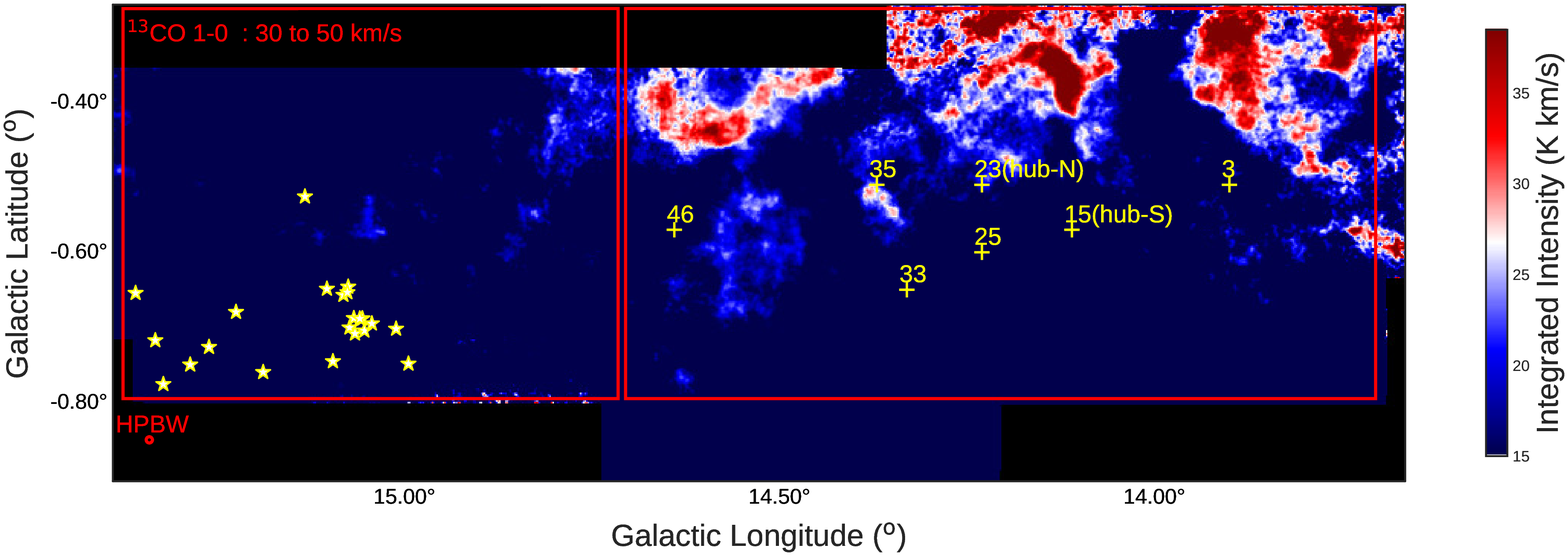}\\
\includegraphics[width=16.cm]{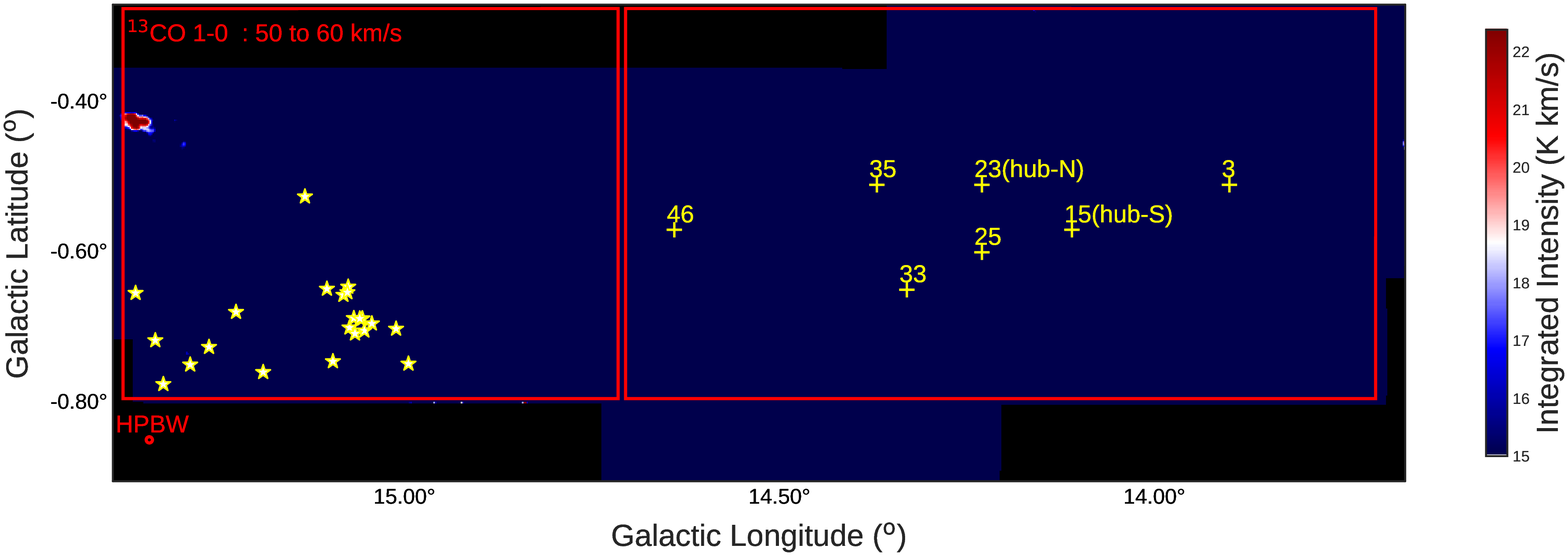}
 \end{center}
\caption{$^{13}$CO\,($J=1-0$) integrated intensity maps of the M17 complex. The velocity range for integration is indicated in the upper-left corner of each panel. 
The red rectangles outline the two prominent star-forming regions: M17-H{\scriptsize II} (left) and M17-IRDC (right). {The star symbols pinpoint the OB star clusters responsible for the giant H{\scriptsize II} region.} The cross symbols pinpoint the locations of the massive cores (M $>$ 500\,\msun) reported by \cite{shimoikura19a}.}
\label{fig:m17_13co_intall}
\end{figure*}

We show the $^{12}$CO\, and $^{13}$CO\, ($J=1-0$) integrated intensities of the observed region integrated over different velocity ranges
in Figures \ref{fig:m17_12co_intall} and \ref{fig:m17_13co_intall}, respectively.

\end{document}